\documentclass[%
 reprint,
 amsmath,amssymb,
 aps,
prd,
intlimits,sumlimits,
showpacs,
]{revtex4-1}

\usepackage{graphicx}
\usepackage{dcolumn}
\usepackage{bm}
\usepackage{comment}

\usepackage{amsfonts, amssymb, wasysym}
\usepackage{mathrsfs} 
\usepackage{graphicx}

\usepackage{caption,subcaption}
\usepackage{amsmath}
\DeclareMathOperator\artanh{arth}
\DeclareMathOperator\sech{sech}

\def\fii{\varphi}
\def\dd{\mathrm{d}}
\def\ii{\mathrm{i}}
\def\pd{\partial}
\def\eps{\varepsilon}
\def\HH{\mathcal{H}}
\def\Real{\mathbb{R}}
\def\Sphere{\mathcal{S}}
\def\DD{\mathcal{D}}
\def\Lie{\pounds}

\def\horeq{\,\dot{=}\,}

\def\NN{\mathcal{N}}
\def\zero{^{\scriptscriptstyle(0)}}

\def\surfkappa{\kappa_{(\ell)}}

\def\KK{\mathcal{K}}
\def\roo{\varrho}

\def\XX{\mathsf{X}}

\renewcommand{\Re}[1]{\mathrm{Re}\, #1}
\renewcommand{\Im}[1]{\mathrm{Im}\, #1}

\def\rr{\mathsf{r}}
\def\vv{\mathsf{v}}
\def\xx{\mathsf{x}}
\def\teta{\vartheta}

\usepackage{marginnote}
\usepackage{color}

\begin{document}

\title[Kerr-Newman as isolated horizon]{Kerr-Newman black hole in the formalism of isolated horizons}

\author{M Scholtz, A Flandera}
\email{scholtz@utf.mff.cuni.cz}\email{flandera.ales@utf.mff.cuni.cz}
\affiliation{Institute of Theoretical Physics, Faculty of Mathematics and Physics, Charles University, \\
  V Hole\v{s}ovi\v{c}k\'ach 2, 182~00, Prague, Czech Republic}

\author{Norman G\"urlebeck}
 \email{norman.guerlebeck@zarm.uni-bremen.de}
\affiliation{%
ZARM, University of Bremen, \\
Am Fallturm, 28359 Bremen, Germany\\
DLR Institute for Space Systems\\
Linzer Str. 1, 28359 Bremen, Germany
}

\begin{abstract}
  The near horizon geometry of general black holes in equilibrium can be conveniently characterized in the formalism of weakly isolated horizons in the form
  of the Bondi-like expansions (Krishnan B, Class.\ Quantum Grav.\ 29, 205006, 2012). While the intrinsic geometry of the Kerr-Newman black hole has been extensively investigated
  in the weakly isolated horizon framework, the off-horizon description in the Bondi-like system employed by Krishnan has not been studied.  We extend Krishnan's work by explicit,
  non-perturbative construction of the Bondi-like tetrad in the full Kerr-Newman spacetime. Namely, we construct the Bondi-like tetrad which is parallelly propagated along
  a nontwisting null geodesic congruence transversal to the horizon and provide all Newman-Penrose scalars associated with this tetrad. This work completes the description
  of the Kerr-Newman spacetime in the formalism of weakly isolated horizons and is a starting point for the investigation of deformed black holes.
\end{abstract}
\pacs{04.20.-q, 04.70.Bw, 04.20.Ex}
\maketitle

\section{Introduction}

The formalism of weakly isolated horizons (WIHs) provides a powerful framework for the analysis of black holes in equilibrium which are a final state of gravitational collapse \cite{Ashtekar1999,Ashtekar-Beetle-Fairhurst-2000,Ashtekar2001,Ashtekar-2002}. One of the earliest applications of the new formalism
was the calculation of statistical mechanical entropy of a black hole in the framework of loop quantum gravity \cite{Ashtekar-Baez-1998}. But WIHs have a plethora of
applications also in classical general relativity (see \cite{AshtekarKrishnanLRR} for a review) and they exhibit a number of properties which make them more realistic in astrophysical settings as compared to standard stationary axisymmetric black holes. For example, WIHs can be embedded in otherwise dynamical spacetimes, they admit the presence of radiation or matter in the neighborhood of the black hole, and they do not require asymptotic flatness. Despite this more general context, WIHs satisfy the usual laws of thermodynamics \cite{Ashtekar-Fairhurst-Krishnan-2000}. Moreover, there is a well defined notion of multipole moments for axially symmetric WIHs \cite{Ashtekar2004}. These moments are intrinsic to the horizon and therefore allow to define
the mass or angular momentum of a black hole quasilocally without the reference to spatial infinity, which is necessary for the definition of Geroch-Hansen multipole moments \cite{Geroch-1970,Hansen-1974}. However, for the higher moments both definitions do not agree in general \cite{Gurlebeck-2014,Gurlebeck-2015} and not even in the case of a Kerr black hole \cite{Ashtekar2004}.

WIHs represent a class of black holes much wider than the standard Kerr-Newman family of solutions \cite{Newman-1965} describing isolated axially symmetric and stationary charged black holes. Nonetheless, the Kerr-Newman metric is the prototypical example of a WIH with well-understood geometry and physical interpretation. One also expects that the geometry of isolated black holes distorted by,
e.g., accreting matter or electromagnetic fields outside the black hole, will deviate only slightly from the Kerr solution; even such small deviations might be measurable in the future experiments \cite{Johannsen-2016,Psaltis-2016}. From the mathematical point of view, one can generate a large class of solutions representing black holes whose intrinsic geometry coincides with the geometry of Kerr-Newman black holes but is (even strongly) distorted
in a neighborhood of the horizon by the appropriate choice of the initial data \cite{Lewandowski-2000}.

The near horizon geometry of a general WIH has been investigated in \cite{Krishnan2012} in the Newman-Penrose formalism. In the neighborhood of the horizon it is possible to introduce
coordinates similar to those used by Bondi \cite{Bondi-1962} in the neighborhood of null infinity and to find a
solution of the field equations near the horizon in a form resembling asymptotic expansions of Newman-Penrose and Newman-Unti near null infinity \cite{Newman-Penrose-1962,Newman-Unti-1962}. An
important feature of the Newman-Penrose formalism is that the field equations naturally split into constraint and evolution equations. Thanks to the properties of WIHs, all constraints can be solved explicitly. Using the evolution part of the equations one can construct an expansion of the solution near the horizon to arbitrary
order. However, in this approach, it is not evident how to choose the initial data in order to reproduce the standard Kerr-Newman solution, nor what the explicit form of the Bondi-like tetrad used in \cite{Krishnan2012} is and what the corresponding Newman-Penrose scalars are.

The formalism of \cite{Krishnan2012} has been recently employed in \cite{Guerlebeck-Scholtz-2017} where we discussed the Meissner effect, i.e.\ the expulsion of electromagnetic fields from
extremal horizons, in the language of WIHs. We have shown that the Meissner effect is an inherent property of extremal, stationary and axisymmetric horizon and takes place also in the strong field regime and independently of the deformations of the black hole. In order to judge, if specific instances of the deformations are physically viable, a description of the full Kerr-Newman spacetime in the formalism of WIHs explicit expressions for Newman-Penrose
quantities in the Bondi-like tetrad is essential.


Whether a given WIH coincides with the horizon of the Kerr metric can be decided using the conditions found in \cite{Lewandowski-Pawlowski-2002}. This result however
includes only the intrinsic geometry of the horizon or the near horizon geometry up to the first order. Bondi-like coordinates
in the Kerr spacetime adapted to the null infinity have been introduced in \cite{Fletcher2003}, but the corresponding tetrad has not been discussed. In \cite{Guerlebeck-Scholtz-2017} we
considered near horizon geometry of the Kerr spacetime up to the first order in Bondi-like coordinates and the Bondi-like tetrad of \cite{Krishnan2012}. 

In this paper we complete the description of the Kerr-Newman black hole in the formalism of WIHs. We perform the construction of \cite{Krishnan2012} explicitly for the Kerr-Newman spacetime. In
particular, we construct a null tetrad adapted to the horizon which is covariantly constant along the nontwisting null congruence transversal to the horizon. We compute all
Newman-Penrose scalars with respect to this tetrad and infer the initial data which has to be given on the horizon and on the transversal null
hypersurface in order to reproduce the Kerr-Newman spacetime. As explained above, this is a starting point for the analysis of physically reasonable deformations of the Kerr black hole.

In Sec.\ \ref{sec:kerr-newman}, we summarize the description of the Kerr-Newman spacetime in standard horizon-penetrating coordinates and Kinnersley null tetrad. The definition of
a WIH and the basic properties of the Bondi-like tetrad employed in \cite{Krishnan2012} are reviewed in Sec.\ \ref{sec:WIH}. The explicit construction is carried out in Sec.\ \ref{sec:bondi-like}.
First, we find the parametrization of nontwisting null geodesic congruences transversal to the horizon. Then, we find
the desired tetrad by a sequence of Lorentz transformations of the Kinnersley one. In order to obtain an explicit solution up to integration, we employ
the coordinate transformation of \cite{Fletcher2003} but adapted to the
horizon rather than to null infinity. In Sec.\ \ref{sec:full-tetrad}, we summarize the results and extract the initial data reproducing the Kerr-Newman spacetime. Definitions of
the Newman-Penrose formalism and, in particular, the transformation properties of the
Newman-Penrose scalars are summarized in App.\ \ref{app:np}. Finally, we visualize and compare the geodesic congruences induced by the Kinnersley tetrad and by the Bondi-like tetrad
in App.\ \ref{sec:visualization}.

\section{Kerr-Newman metric}
\label{sec:kerr-newman}
The main purpose of this section is to set up the notation and conventions used in this paper and to provide equations for later references. Typically, we employ the abstract
index notation \cite{PenroseRindlerI} and denote the abstract indices by Latin letters from the beginning of the alphabet, $a, b, \dots$. Greek indices will take values from 0 to 3
and they will denote components of a tensor with respect to particular coordinates. Indices $I, J, \dots $ will take values $2,3$. We use the signature $(+---)$ so that
the Newman-Penrose (NP) null tetrad \cite{Newman-Penrose-1962,Stewart1993} is normalized by the conditions $\ell^a n_a = 1, m^a\,\bar{m}_a = -1$. For a review of the relevant definitions and relations of the NP formalism, see Appendix \ref{app:np}. The Riemann tensor is defined by $2\nabla_{[c}\nabla_{d]}X^a = - R^a_{\phantom{a}bcd}X^b$, where square brackets denote the total antisymmetrization. In the NP formalism, Einstein's equations for electro-vacuum read $\Phi_{mn} = \phi_m \bar{\phi}_n$, $\Lambda = 0$.

We start with the Kerr-Newman metric describing a black hole of mass $M$, spin $a$ and charge $Q$ in the ingoing null coordinates $x^\mu=(v,r,\theta,\fii)$, in which the line element
takes the form \cite{Newman-1965,Krishnan2014} 
\begin{widetext}
\begin{align}\label{eq:KN metric}
 \dd s^2 &= \left( 1-\frac{2\,M\,r-Q^2}{|\rho|^2} \right)\dd v^2 - 2\,\dd v\,\dd r + \frac{2\,a}{|\rho|^2}\left( 2\,M\,r-Q^2 \right)\sin^2\theta\,\dd v \,\dd \fii  \nonumber\\
&\qquad{}+ 2\,a\sin^2\theta\,\dd r\,\dd \fii  - |\rho|^2\dd\theta^2 + \frac{\sin^2\theta}{|\rho|^2}\left( \tilde{\Delta}\,a^2\sin^2\theta - \left( a^2+r^2 \right)^2 \right)\dd\fii^2,
\end{align}
\end{widetext}
where the functions $\rho$ and $\tilde{\Delta}$ are given by relations
\begin{align}\label{eq:rho Delta def}
\rho &= r + \ii\,a\,\cos\theta,  &
 \tilde{\Delta} &= a^2 +r^2 - 2\,M\,r + Q^2.
\end{align}
The outer and inner horizons are located at
\begin{align}\label{eq:r horizons}
r_\pm &= M \pm \sqrt{M^2-a^2 - Q^2},
\end{align}
respectively.
We will write $X \horeq Y$ when the two quantities are equal on the outer horizon $r_+$, for example, $\tilde{\Delta}\horeq 0$. 

In order to analyze the Kerr-Newman metric in the NP formalism, we first introduce the standard Kinnersley null tetrad \cite{Kinnersley-1969,Krishnan2014} adapted to the principal null directions of the
Kerr-Newman metric,
\begin{align}
\ell_K  &=  \pd_v + \frac{\tilde{\Delta}}{2\!\left( a^2+r^2 \right)}\,\pd_r + \frac{a}{a^2+r^2}\,\pd_\fii, \nonumber\\
n_K&= - \frac{a^2+r^2}{|\rho|^2}\pd_r, \label{eq:kinnersley}\\
m_K &= \frac{1}{\sqrt{2}\,\rho}\!\left( \ii\,a\sin\theta\,\pd_v + \pd_\theta + \frac{\ii}{\sin\theta}\,\pd_\fii \right)\!,\nonumber
\end{align}
where the fourth vector of the tetrad $\bar{m}_K^a$ is the complex conjugate of $m_K^a$. Notice also that the triad $(\ell_K^a, m_K^a, \bar{m}_K^a)$ is tangent to the horizon. The spin
coefficients associated with the tetrad \eqref{eq:kinnersley} are
\begin{align}
\kappa_K &= \sigma_K = \nu_K = \lambda_K = 0, & \gamma_K &=  - \frac{a\left( a+\ii\,r\cos\theta \right)}{\rho\,\bar{\rho}{}^{\,2}}, \nonumber\\
 \roo_K &= - \frac{\tilde{\Delta}}{2\,\bar{\rho}\left( a^2+r^2 \right)},&  
 \tau_K & = -\frac{\ii\,a \sin\theta}{\sqrt{2}|\rho|^2}, \nonumber\\
 \eps_K &= - \frac{M\,a^2 + r\,Q^2-M\,r^2}{2(a^2+r^2)^2}, &
 \mu_K &= - \frac{a^2+r^2}{\rho\,\bar{\rho}{}^{\,2}}, \nonumber\\
  \pi_K &= \frac{\ii\,a\sin\theta}{\sqrt{2}\,\bar{\rho}{}^{\,2}}, &
                                                                                         a_K &= \frac{\ii\,a - r\cos\theta}{\sqrt{2}\,\bar{\rho}{}^{\,2} \sin\theta}\,,
 \label{eq:spin coefficients 1}
\end{align}
where we have denoted $a_K = \alpha_K - \bar{\beta}_K$; in addition, $\pi_K = \alpha_K + \bar{\beta}_K$. These relations can be geometrically
interpreted as follows, cf.\ App.\ \ref{app:np} and \ref{sec:visualization}. The vector $\ell_K^a$ is tangent to a congruence of null curves, which are geodesics ($\kappa_K=0$) and
shear-free ($\sigma_K=0$). The expansion and twist of $\ell_K^a$ vanish on the horizon ($\roo_K\horeq 0$). Similarly, $n_K^a$ is tangent to null geodesics ($\nu_K=0$), which are
shear-free ($\lambda_K = 0$), but have non-vanishing expansion ($\Re\mu_K$) and twist ($\Im\mu_K$). Thus, the vector field $n_K^a$ is not hypersurface orthogonal. In fact, vectors
$\ell_K^a$ and $n_K^a$ are, at each point, the principal null directions of the Weyl tensor, so that the only non-vanishing Weyl scalar is
\begin{align}\label{eq:Psi2}
\Psi^K_2 &= - \frac{M}{\bar{\rho}{}^{\,3}} + \frac{Q^2}{\bar{\rho}{}^{\,3}\,\rho}. 
\end{align}
In the presence of charge, the non-vanishing component of the trace-free part of the Ricci tensor is
\begin{align}\label{eq:Phi11}
 \Phi^K_{11} &= \frac{Q^2}{2|\rho|^4},
\end{align}
while the scalar curvature $\Lambda$ vanishes. Comparing the NP form of Einstein's equation $\Phi^K_{11}=|\phi^K_1|^2$ and \eqref{eq:Phi11}, and using the Maxwell equations, we find
\begin{align}\label{eq:phi1}
 \phi^K_1 &= \frac{Q}{\sqrt{2}\,\bar{\rho}^{2}},
\end{align}
while $\phi^K_0$ and $\phi^K_2$ vanish. Hence, the principal null directions of $F_{ab}$ are aligned with $\ell_K^a$ and $n_K^a$. The four-potential of the electromagnetic field turns out to be
\begin{align}
 A_\mu \dd x^\mu &= \frac{\sqrt{2}\,Q \,r}{|\rho|^2}\!\left( \dd v - a\sin^2\theta\,\dd\fii \right)\!.
\end{align}
Finally, the Kerr-Newman metric admits a Killing-Yano form \cite{Jezierski2006,Kalnins-1989} which, in the Kinnersley tetrad \eqref{eq:kinnersley}, reads
\begin{align}
  Y^{ab} &= -2\, a\cos\theta\; \ell^{[a}_K \, n_K^{b]} + 2\,\ii\,r\,m^{[a}_K\, \bar{m}_K^{b]},
\end{align}
or, in coordinates,
\begin{align}\label{eq:Killing-Yano tensor}
 Y &=  \dd \fii\wedge\left[ a^2\cos\theta\sin^2\theta\,\dd r - r\left( a^2+r^2 \right)\sin\theta\,\dd \theta\right] \nonumber\\
 &\qquad{}+ \dd v \wedge\left[-a\cos\theta \,\dd r + a\,r\sin\theta\,\dd \theta\right]\!.
\end{align}
For any geodesic vector $X^a$, the one-form $k_a=Y_{ab} X^b$ is covariantly constant along $X^a$.

\section{Weakly isolated horizons}
\label{sec:WIH}

A spacetime $M$ is said to admit a \emph{nonexpanding horizon} \cite{Ashtekar-2002}, if it contains a null hypersurface $\HH\subset M$ with the topology $\Real\times\Sphere^2$ on
which Einstein's equations hold and the energy-momentum tensor $T_{ab}$ satisfies the ener\-gy condition that $T_{ab} \ell^b$ is causal (i.e., timelike or null) and future pointing for
any future null vector $\ell_a$ normal to $\HH$. Moreover, any such normal is nonexpanding. It turns out that the spacetime connection $\nabla_a$ induces a preferred
connection $\DD_a$ on $\HH$ and gives rise to a \emph{rotational 1-form} $\omega_a$ defined by
\begin{align}\label{eq:rotational 1-form}
\DD_a \ell^b &\horeq \omega_a \ell^b.
\end{align}
Since the choice of the null normal $\ell^a$ is not unique, it is natural to fix it by the requirement \cite{Ashtekar2001}
\begin{align}\label{eq:WIH condition}
 [\Lie_\ell, \DD_a] \ell^b \horeq 0
\end{align}
which is equivalent to $\Lie_\ell \omega_a \horeq 0$; here $\Lie_\ell$ is the Lie derivative along $\ell^a$. This leads to a definition of a \emph{weakly isolated horizon} (WIH) as
a nonexpanding horizon $\HH$ equipped with an equivalence class of null normals $[\ell^a]$, where elements of $[\ell^a]$ differ just by constant rescaling, satisfying the
condition \eqref{eq:WIH condition}. This condition guarantees that the zeroth law of black hole thermodynamics is satisfied on $\HH$ and that the pull-back of $\ell^a$ is a
Killing vector of the induced degenerate metric on $\HH$.

Using the geometrical properties of the WIHs it is possible to construct Bondi-like coordinates and a null, Bondi-like, tetrad adapted to these coordinates. Details of
this construction and perturbative solutions of the Einstein-Maxwell equations near a WIH were given in \cite{Krishnan2012}. Here, we briefly review the main steps of the construction.

By definition, a given WIH $(\HH, [\ell^a])$ has a preferred foliation by topological spheres. A coordinate $\vv$ on $\HH$ is defined by the requirement that it is constant
on each sphere $\Sphere_\vv$ of the foliation and, in addition, $D \vv \horeq 1$, where $D = \ell^a \nabla_a$. Next, arbitrary
coordinates $\xx^I$, $I=2,3$, are introduced on the sphere $\Sphere_0$ along with two arbitrary complex null vectors $m^a$ and $\bar{m}^a$ tangent to $\Sphere_0$. To get a basis of
the space tangent to $\HH$, the vector $m^a$ is propagated along $\ell^a$ by the condition
\footnote{Notice that, thanks to the properties of nonexpanding horizon, Lie dragging preserves the normalization of the vectors.}
\begin{align}
 \Lie_\ell m^a &\horeq 0.
\end{align}
Similarly, coordinates $\xx^I$ are propagated off the sphere $\Sphere_0$ along $\ell^a$ by the condition $D\xx^I \horeq 0$ so that we have a coordinate system $(\vv, \xx^2, \xx^3)$ for
the entire horizon $\HH$. The triad $(\ell^a, m^a, \bar{m}^a)$ can now be completed to a full NP tetrad $(\ell^a, n^a, m^a, \bar{m}^a)$ on the horizon. In order to
obtain the tetrad in the neighbourhood of the horizon, the vector field $n^a$ on $\HH$ is extended geodesically off the horizon and the remaining vectors are parallely
transported along the resulting geodesic congruence, i.e., all vectors are propagated off the horizon by the conditions
\begin{align}
 \Delta n^a &= \Delta \ell^a = \Delta m^a = 0,
\end{align}
where $\Delta = n^a \nabla_a$. The coordinate $\rr$ is defined as an affine parameter along the geodesics $n^a$ and the coordinates $\vv$ and $\xx^I$ are extended
off the horizon by the conditions $\Delta \vv = \Delta \xx^I = 0$. In this way, the NP tetrad and the coordinates $\xx^\mu=(\vv,\rr,\xx^2,\xx^3)$ are introduced on the
horizon and in its neighborhood. In these coordinates, the vectors of the null tetrad read
\begin{equation}
    \begin{aligned}
    \ell &= \pd_{\vv} + U\,\pd_{\rr} + X^{{I}}\,\pd_{{I}}, &
    n &= -\, \pd_{\rr}, & \\
    m &= \Omega\, \pd_{\rr} + \xi{}^{{I}} \pd_{{I}}.\label{eq:Krishnan tetrad}
    \end{aligned}
\end{equation}
By construction, the following functions vanish on $\HH$:
\begin{align}\label{eq:frame on H}
 U &\horeq X^I \horeq  \Omega \horeq 0.
\end{align}
Throughout the paper, we reserve the symbol $\xx^\mu$ for the coordinates introduced in this section. Adopting the terminology of \cite{Fletcher2003}, we will refer to
the coordinates $\xx^\mu$ and the tetrad (\ref{eq:Krishnan tetrad}) as the ``Bondi-like coordinates'' and the ``Bondi-like tetrad'', respectively. 

By construction of the tetrad,
\begin{align}
  \gamma & = \nu = \tau=0, &
                           \pi = \alpha + \bar{\beta},
\end{align}
everywhere, and $\kappa \horeq 0$ on the horizon. Moreover,
\begin{align}
  \mu &= \bar{\mu},
\end{align}
which means that $n^a$ is twist-free and, hence, orthogonal to hypersurfaces $\NN_{\vv}$ of constant $\vv$, which are transversal to the horizon and intersect it in the spherical cuts $\Sphere_{\vv}$. Since the normal to the horizon $\ell^a$ is by assumption nonexpanding and orthogonal to $\HH$, we have
\begin{align}
  \roo &\horeq 0,
\end{align}
which, together with the energy condition imposed on the energy-momentum tensor and the Ricci identities in the NP formalism, implies
\begin{align}
  \sigma & \horeq 0, &
                       \Psi_0 &\horeq \Psi_1 \horeq 0, &
                                                        \phi_0 & \horeq 0,
\end{align}
i.e., the horizon is also shear-free and there is no gravitational or electromagnetic radiation crossing the horizon.

Having established the null tetrad and the coordinate system, it is possible to solve the Einstein equations perturbatively in the neighbourhood of $\HH$. More precisely, we regard the
spacetime as a solution to a characteristic initial value problem with the initial data given on the horizon $\HH$ and any null hypersurface, say $\NN_0$, intersecting the horizon
(see \cite{Racz-2007,Racz-2014} for the precise formulation and existence results).

Adopting the notation of \cite{Krishnan2012,Ashtekar-2002}, the initial data on $\Sphere_0$ consists of the following NP scalars:
\begin{align}
 \Sphere_0: \quad  \pi\zero, \phi_1\zero, \mu\zero, \lambda\zero, \xi^I|_{\Sphere_0}, \surfkappa,
\end{align}
where the $\xi^I|_{\Sphere_i}$ are the components of $m^a$ on the sphere $\Sphere_0$ which, in turn, define the two-dimensional metric on $\Sphere_0$. By the properties of WIHs, the surface gravity
of the normal $\ell^a$,
\begin{align}
 \surfkappa &\horeq \eps\zero + \bar{\eps}\zero,
\end{align}
is constant over the horizon -- the zeroth law of thermodynamics. The quantity $\phi_1\zero$ is the NP component of the electromagnetic field and its real and imaginary parts describe the electric as well as magnetic flux density through $\Sphere_0$, respectively. From the Ricci identities in the NP formalism one
can then calculate the quantities
\begin{align}
 a\zero = \alpha\zero - \bar{\beta}\zero,  \Psi_2\zero ~\text{and}~\Psi_3\zero
\end{align}
on $\Sphere_0$. In the case of axisymmetric, stationary WIHs, the Weyl scalar $\Psi_2\zero$ encodes the horizon multipole moments \cite{Ashtekar2004}. The quantity
$a\zero$ defines the connection on $\Sphere_0$. In addition, the Ricci identities determines the evolution of all these quantities along the horizon and, as it turns out, only the spin coefficients
$\mu$ and $\lambda$ and the Weyl scalar $\Psi_3$ depend on the coordinate $\vv$ on $\HH$.

\begin{figure}
\begin{center}
 \includegraphics[width=0.4\textwidth]{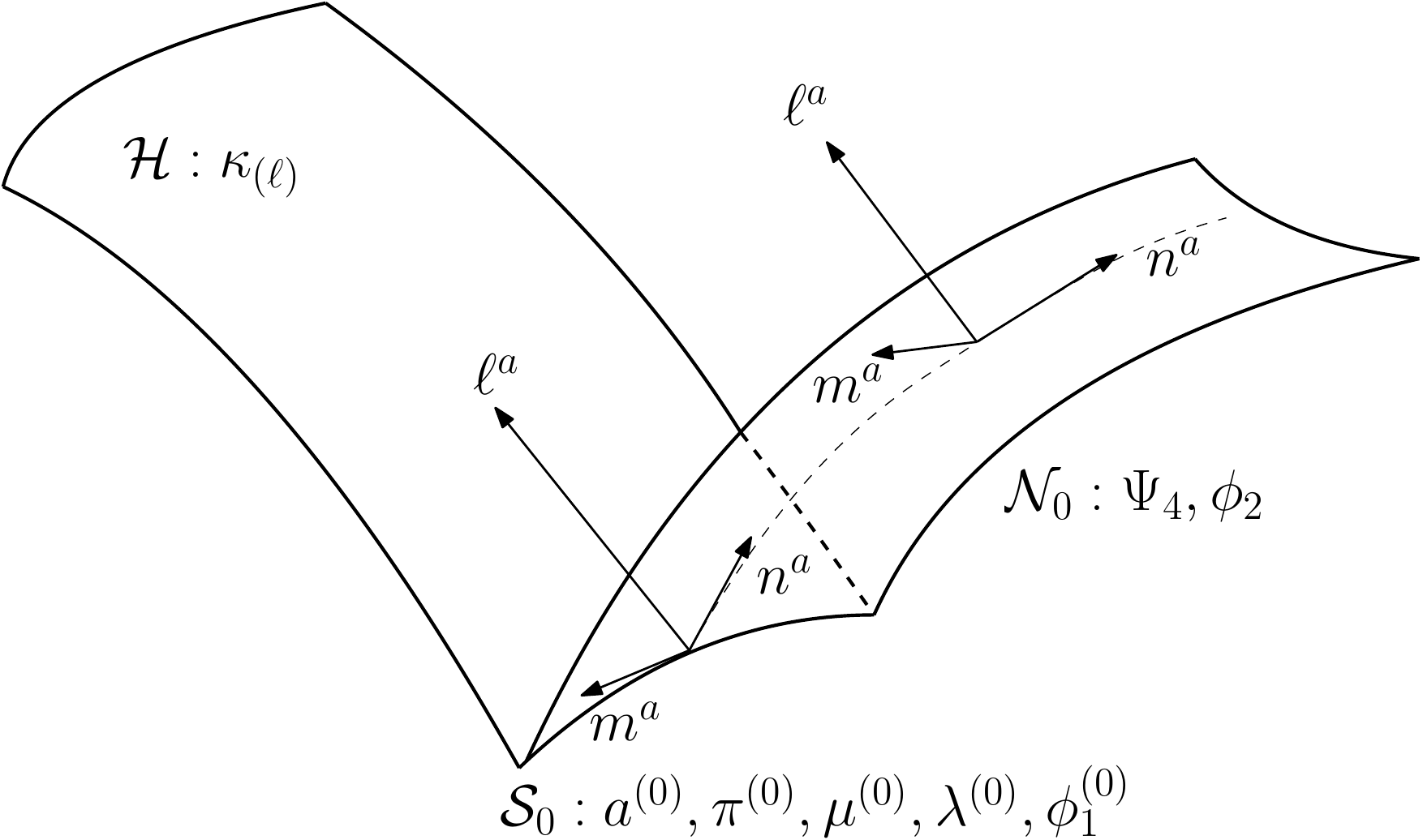}
 \end{center}
 \caption{Characteristic initial value problem for WIH.}
 \label{fig:tetrad}
\end{figure}

The two remaining NP quantities, the Weyl scalar $\Psi_4$ and the component of the electromagnetic field $\phi_2$ can be specified freely on the transversal null hypersurface $\NN_0$. The Ricci and
Bianchi identities as well as the Maxwell equations in the NP formalism then determine the solution of the full Einstein-Maxwell equations in the neighborhood of $\HH$. The geometrical
setup is schematically sketched in Figure \ref{fig:tetrad}. 

With this formulation of the initial value problem, one can expand all geometrical quantities into a series in the coordinate $\rr$ and find
a perturbative solution of the Einstein-Maxwell equations in a neighbourhood of $\HH$. This was done in \cite{Krishnan2012}. The question
now arises, how to choose the initial data on $\HH$ and $\NN_0$ in order to reproduce any particular spacetime. For a Schwarzschild spacetime, this is a trivial
task because the principal null directions of the Weyl tensor are already nontwisting and, therefore, the Kinnersley tetrad reduces to the desired Bondi-like one for $a=0$. We will
show that the physically more interesting solution, the Kerr-Newman spacetime, also allows the analytic construction of the Bondi-like tetrad.

\section{Bondi-like tetrad for Kerr-Newman spacetime}
\label{sec:bondi-like}

\subsection{Nontwisting null geodesics}
Let $(v,r,\theta,\fii)$ be the standard ingoing null coordinates introduced in Sec.\ \ref{sec:kerr-newman}. In order to construct a null tetrad which meets
the criteria imposed in \cite{Krishnan2012} for the Kerr-Newman spacetime, we first construct a nontwisting null geodesic congruence with the tangent vector $n_B^a$
which will later be identified with the vector of the Bondi-like null tetrad. Using the well-known se\-parability of the geodesic equation on the Kerr-Newman
background \cite{Carter1968}, a generic null geodesic can be characte\-rized by three constants $E, L$ and $\KK$, in terms of which the components of $n_B^a$ read
\begin{equation}
\begin{aligned}\label{eq:xmu dot}
n_B^v  &= - \frac{1}{|\rho|^2} \!\left( a^2\,E\sin^2\theta + a\,L + \frac{a^2+r^2}{\tilde{\Delta}}\!\left( \sqrt{R } - P \right)\!  \right)\!, \\
n_B^r  &= - \frac{1}{|\rho|^2}\,\sqrt{R}\,, \\
n_B^\theta &= - \frac{1}{|\rho|^2}\,\sqrt{\Theta}\,, \\
n_B^\fii  &= - \frac{1}{|\rho|^2} \!\left( a\,E + L\,\sin^{-2}\theta + \frac{a}{\tilde{\Delta}} \!\left( \sqrt{R }-P \right)\! \right)\!,
\end{aligned}
\end{equation}
where functions $\Theta,\, P$ and $R $ are defined by
\begin{equation}
\begin{aligned}
 \Theta &= \KK - \left( L+a\,E \right)^2+\left(a^2\,E^2-L^{2}\sin^{-2}\theta\right)\cos^2\theta, \\
 P &= a\,L + E\left( a^2+r^2 \right), \qquad R  = P^2 - \KK\, \tilde{\Delta}.
\end{aligned}
\end{equation}
The constants of motion $E$ and $L$ represent the energy and the angular momentum of the geodesic and $\KK$ is the Carter constant which arises from the separation of the Hamilton-Jacobi
equation or from the projection of the Killing tensor of the Kerr-Newman spacetime \cite{WalkerPenrose1970}. 

In the next step, we promote \eqref{eq:xmu dot} to a geodesic \emph{congruence}, where each geodesic of the congruence is parametrized by possibly different values of $E, L$ and $\KK$. In
this way, these parameters become functions of the position, assigning the corresponding values to a geodesic passing through a given point. We wish to choose the functions $E, L$ and $\KK$ in such a
way that the resulting congruence is nontwisting. In order to accomplish that, we require that the covariant vector $(n_B)_a$ be a gradient, i.e.,
\begin{align}\label{eq:dV}
 (n_B)_\mu \dd x^\mu \equiv E\,\dd v - \frac{a^2}{P+\sqrt{R}}\,\dd r + \sqrt{\Theta}\,\dd\theta + L\,\dd \fii = \dd \vv
\end{align}
for some function $\vv$. Since $\pd_v$ and $\pd_\fii$ are the Killing vectors of the Kerr-Newman metric, we assume $E = E(r,\theta)$ and
similarly for $L$ and $\KK$. An inspection of the integrability condition for the existence of a function $\vv$ in  Eq.\ \eqref{eq:dV}
\begin{align}\label{eq:EL constant}
(\dd n_B)_\mu \wedge \dd x^\mu &= 0,
\end{align}
then shows that $E$ and $L$ must be constant everywhere, while $\KK$ must satisfy the condition
\begin{align}\label{eq:K}
\sqrt{\Theta}\,\frac{\pd \KK}{\pd \theta} &= - \sqrt{R}\,\frac{\pd \KK}{\pd r},
\end{align}
which is equivalent to $n_B^a \nabla_a\KK=0$ and does not impose a new condition, since $\KK$ is the constant of motion. 

Because $L$ must be a constant, the components of the geodesic \eqref{eq:xmu dot} become singular on the axis $\theta=0$. In fact, this is a singularity of a congruence rather than coordinate
singularity, since, for example, the expansion $\nabla_a n_B^a$ diverges there. In order to avoid the singular behavior, we set $L=0$. Notice that although it is natural to expect that nontwisting congruence
has zero angular momentum, vanishing of the twist alone is compatible with any value of $L$.

Then, imposing that the congruence be symmetric under
reflection across the equatorial plane \footnote{That is, radial geodesics emanating from the equatorial plane will remain in this plane.}, we have to choose $\KK = a^2\,E^2$. Finally,
similarly to \cite{Fletcher2003}, we also choose $E=1$ for convenience. Hence, we set
\begin{align}
 L &= 0, & \KK &= a^2,
\end{align}
so that the congruence \eqref{eq:xmu dot} simplifies to
\begin{subequations}
\begin{align}
 n_B^v &= - \frac{a^2}{|\rho|^2}\left(\sin^2\theta- \frac{P}{P+\sqrt{R}} \right)\!, \\
 n_B^r &= - \frac{1}{|\rho|^2} \, \sqrt{R}\,, \\
 n_B^\theta &= - \frac{a\cos\theta}{|\rho|^2}\,, \\
 n_B^\fii &= - \frac{a}{|\rho|^2}\left(  1 - \frac{a^2}{P+\sqrt{R}}\right)\!,
\end{align}
\label{eq:n congruence}
\end{subequations}
where functions $P$ and $R$ now read
\begin{align}\label{eq:Theta P R def}
  P &= a^2 + r^2, &
                    R &=  r^4+a^2\,r^2+2\,a^2\,M\,r-a^2\,Q^2.
\end{align}

\subsection{Lorentz transformations}
\label{sec:lorentz}
Having found an appropriate null congruence (\ref{eq:n congruence}), we wish to complete $n_B^a$ to a Bondi-like null tetrad, i.e.\ find vectors $\ell_B^a$ and $m_B^a$ which are
covariantly constant along $n_B^a$. The standard technique to parallely propagate a frame along null geodesics using the Killing-Yano form (\ref{eq:Killing-Yano tensor}) was
developed in \cite{Marck-1983} and recently generalized to $n$-dimensional spacetimes admitting a conformal Killing-Yano form in \cite{Kubiznak-2009}. The vector $k_a = Y_{ab}n_B^b$,
where $n_B^b$ is given by (\ref{eq:n congruence}) is parallelly propagated along $n_B^a$ and it is spacelike rather than null. Still, one can find another parallelly propagated spacelike vector
$\tilde{k}^a$ and take the complex linear combination of $k^a$ and $\tilde{k}^a$ in order to form the null vector $m_B^a$ which is covariantly constant along $n_B^a$, and finally
complete the tetrad with vector $\ell_B^a$, which is then automatically covariantly constant too. Unfortunately, the frame obtained in this way does not
satisfy the conditions imposed on the Bondi-like tetrad, i.e.\ the vector $m_B^a$ is not tangent to the horizon and $\ell_B^a$ is not a normal to the horizon. Thus, one
has to rotate the frame on the horizon to a desired Bondi-like tetrad with an appropriate matrix $R$ and then require the matrix to be constant along $n_B^a$. However,
the outlined procedure requires explicit solution of the geodesic equation at two stages: finding vector $\tilde{k}^a$ and propagating the matrix $R$ off the horizon. Such explicit solution
can be found in terms of the elliptic integrals of the first kind but one has to solve the polynomial equation of the fourth order. 

Because of these complications with the standard methods, we adopt a different approach in which we do not need the Killing-Yano form at all. In our approach we also get
integrals which cannot be calculated explicitly or the resulting formulas are too complex to be included in the paper. However, we present the tetrad in the form suitable
both for symbolic manipulations and numerical calculations. We perform a sequence of Lorentz
transformations which rotate the initial Kinnersley tetrad (\ref{eq:kinnersley}) to a Bondi-like tetrad. Using the boost and null rotation about $\ell_K^a$,
we rotate the vector $n_K^a$ to the direction (\ref{eq:n congruence}), then by a spin and null rotation about $n_B^a$ we eliminate the spin coefficients $\gamma$ and $\tau$,
which yields the triad $(\ell_B, m_B, \bar{m}_B)$  tangent to the horizon and parallelly propagated along $n_B^a$. By this method we get not only the tetrad, but
also the spin coefficients and the Weyl and Maxwell scalars, because it is easy to transform the spin coefficients step by step but very difficult to
calculate them directly from the resulting tetrad, even with the help of computer algebra systems. 

\paragraph{Boost.} In the first step, we perform a boost (\ref{np:boost}) in the plane spanned by $\ell_K^a$ and $n_K^a$ of the Kinnersley tetrad (\ref{eq:kinnersley}) with the parameter
\begin{align}
  A^2 &= \frac{2\,P}{P+\sqrt{R}},
\end{align}
obtaining a new tetrad $(\ell_1,n_1,m_1,\bar{m}_1)$ and corresponding spin coefficients.

\paragraph{Null rotation about $\ell^a$.} Next, we rotate the tetrad about $\ell_1^a$ to a new tetrad $(\ell_2,n_2,m_2,\bar m_2)$ such that the vector field $n^a_2$ coincides with the nontwisting vector field (\ref{eq:n congruence}). We choose the parameter $c$ of the null rotation (\ref{np:rotation l}) to be
\begin{align}
  c &= - \frac{a\,e^{-\ii\,\theta}}{\sqrt{2}\,\bar{\rho}}.
\end{align}
Then, $n_2$ is given by (\ref{eq:n congruence}).
\paragraph{Spin.} At this stage, the vector $n_2^a$ is tangent to the desired nontwisting, affinely parametrized congruence of geodesics, i.e.\ $\Delta_2 n_2^a = 0$ with the respective radial derivative $\Delta_2=n^a_2\nabla_a$. Moreover, the
triad $(\ell_2^a, m_2^a, \bar{m}_2^a)$ is tangent to the horizon, where $\ell_2^a$ is a generator of the horizon satisfying \eqref{eq:WIH condition}. However, this triad is not covariantly constant
along $n_2^a$, since we have (cf.\ the transport equations (\ref{np:transport D l Delta n}))
\begin{subequations}
  \begin{align}
    \Delta_2 \ell_2^a &= - \bar{\tau}_2\,m_2^a - \tau_2\,\bar{m}_2^a, \\
    \Delta_2 m_2^a &= - \tau_2\,n_2^a + (\gamma_2 - \bar{\gamma}_2)\,m_2^a.
  \end{align}
\end{subequations}
The coefficient $\gamma_2$ which is now purely imaginary ($n_2^a$ is affinely parametrized) can be eliminated completely by the spin (\ref{np:spin}) with the parameter $\chi = -\theta/2$, i.e.\
$m^a_3 = e^{-\ii\,\theta}\,m_2^a$. The nonvanishing spin coefficients now are
\begin{subequations}\label{eq:spin coefficients last}
  \begin{align}    
    \tau_3 &= \frac{a}{\sqrt{2}\,|\rho|^2}\!\left( \frac{\tilde{\Delta}}{P+\sqrt{R}} - \ii\,e^{-\ii\,\theta}\sin\theta\right)\!,\\
    \roo_3 &= - \frac{\tilde{\Delta}}{\bar{\rho}(P+\sqrt{R})},\\
    \eps_3 &= \frac{r}{a^2} - \frac{1}{2\sqrt{R}} \! \left( \frac{2\,r^3}{a^2}+M+r\right)\!, \\
    \pi_3 &= \frac{P-\sqrt{R}}{\sqrt{2}\,a\,\bar{\rho}^2} - \frac{\sqrt{2}\,a}{\bar{\rho}} \! \left(\frac{r}{a^2} - \frac{M+r+2\,r^3/a^2}{2\sqrt{R}}\right) \nonumber \\
        &\qquad{}+ \left(\frac{2\,\ii\,a - \rho\cos\theta}{2\sqrt{2}\,\bar{\rho}^2\sin\theta} + \frac{\cot\theta}{2\sqrt{2}\,\bar{\rho}}\right)e^{\ii\,\theta}, \\
    \alpha_3 &= - \frac{a}{\sqrt{2}\,\bar{\rho}} \! \left(\frac{r}{a^2}- \frac{M+r+2\,r^3/a^2}{2\sqrt{R}} - \frac{P-\sqrt{R}}{a^2\,\bar{\rho}}\right) \nonumber \\
        &\qquad{}+ \frac{e^{\ii\,\theta}}{2\sqrt{2}\,\bar{\rho}}\!\left(\frac{2\,\ii\,a - \rho \cos\theta}{\bar{\rho} \sin\theta} - \ii\right)\!.
  \end{align}
  \begin{align}
      \beta_3 &=\frac{1}{2\, \sqrt{2} \,\rho } \bigg[e^{-\ii\,\theta}(\cot \theta-\ii) \nonumber \\
        &\qquad{}-2 \,a  \!\left( \frac{r}{a^2} - \frac{M+r+2\,r^3/a^2}{2\,\sqrt{R}}\right)\! \bigg]\!, \\
      \mu_3 &= \frac{1}{2\sqrt{R}\,|\rho|^2} \Big(-2 \,r^3   -a^2(M+r) \nonumber \\
            &\qquad{}- a\,\sqrt{R}\cos2\theta\,\csc\theta\Big)\!, \\
      \lambda_3 &= \frac{a}{2\sqrt{R}\,\bar{\rho}^3}\Big( - 2 a Q^2 + a \,r(3\,M+r) \nonumber \\
        &\qquad{}+ \ii\, (2\,r^3+a^2(M+r))\cos\theta \nonumber \\
        &\qquad{}+ \sqrt{R}\,(r\cos 2\theta - \ii\,a\cos\theta)\csc\theta  \Big)\!.
  \end{align}
\end{subequations}

By now, the only freedom in the choice of the tetrad is the null rotation about $n^a$. In order to preserve the property that the triad $(\ell_3^a, m_3^a, \bar{m}_3^a)$ be tangent
to the horizon, the parameter $d$ of null rotation (\ref{np:rotation n}) must vanish on the horizon. The spin coefficient $\tau$ transforms, according to (\ref{np:rotation n spins}),
by $ \tau_4 = \tau_3 - \Delta_3 d$, where $\Delta_3=n_3^a\nabla_a$. Thus, in order to eliminate $\tau_3$ we have to solve the equation
\begin{align}\label{eq:Delta d}
  \Delta_3 d &= \tau_3, \qquad d \horeq 0.
\end{align}
It is exactly the initial condition $d \horeq 0$ which makes the problem difficult, otherwise the equation could be easily solved with the ansatz $d = f(r) + g(\theta)$. In order
to implement the initial condition, we have to employ a coordinate transformation which will eliminate the non-radial components of $n_3^a$.

\newpage

\paragraph{Bondi-like coordinates.} \label{sec:Bondi-coords} Having satisfied the integrability conditions \eqref{eq:EL constant} for Eq.\ \eqref{eq:dV}, we can employ $\vv$ as a
new coordinate, eliminating the $\vv$-component of $n_3^a$. The angular coordinates which are constant along $n_3^a$ can be conveniently introduced following the procedure of
\cite{Fletcher2003}. Hence, we define the new coordinates $\vv$, $\teta$ and $\tilde{\phi}$ by
\begin{subequations}
  \begin{align}
    \vv &= v - \int_{r_+}^r\frac{a^2\,\dd r}{P+\sqrt{R}} +a \sin\theta,\label{eq:v-transf} \\
    \sin\theta &= \tanh\XX,\label{eq:theta-transf} \\
    \tilde{\phi} &= \fii + J(r),\label{eq:phi-transf}
  \end{align}
\end{subequations}
where
\begin{subequations}
  \begin{align}
    \XX &= \alpha(r) + \artanh \sin\teta,\label{eq:X-def} \\
    \alpha(r) &= \int_{r_+}^r \frac{a\,\dd u}{\sqrt{u^4 +a^2\,u^2+2\,a^2\,M\,u - a^2\,Q^2}}, \label{eq:alpha-def}\\
    J(r) &= -\int_{r_+}^r\frac{a}{P(u)+\sqrt{R(u)}} \! \left( 1 + \frac{u^2}{\sqrt{R(u)}} \right) \dd u.
  \end{align}
\end{subequations}

\bigskip

With these choices of $\alpha$ and $J$, coordinates $(\teta, \tilde{\phi})$ coincide with the coordinates $(\theta,\fii)$ on the horizon (in \cite{Fletcher2003} they coincide at infinity),
and equatorial plane is given everywhere by $\theta = \teta = \pi/2$. Moreover, $\teta$ and $\tilde{\phi}$ are constant along $n_3^a$. Using the relations
\begin{subequations}
  \begin{align}
    \dd v &= \dd \vv + \left( \frac{1}{P+\sqrt{R}} - \frac{1}{\sqrt{R}\,\cosh^2\XX}\right)a^2\,\dd r \nonumber \\
          &\qquad{} - \frac{a\,\dd\teta}{\cosh^2 \XX\,  \cos\teta}, \\
    \dd \theta &= \frac{1}{\cosh\XX}\!\left( \frac{a\,\dd r}{\sqrt{R}} + \frac{\dd \teta}{\cos\teta}\right)\!, \\
    \dd\fii &= \frac{a}{\sqrt{R}} \!\left( 1 - \frac{a^2}{P+\sqrt{R}}\right)\! \dd r + \dd \tilde{\phi},
  \end{align}
\end{subequations}
one can deduce the form of the metric tensor in these coordinates:
\begin{widetext}
  \begin{align}
    \dd s^2 &= \left(1+\frac{Q^2-2\,M\,r}{|\rho|^2}\right)\!\left( \dd \vv - \frac{2\,a}{\cosh^2\XX\,\cos\teta} \dd\teta \right)\dd \vv
              - \frac{2|\rho|^2}{\sqrt{R}}\,\dd \vv\,\dd r - \frac{2\,a}{|\rho|^2}\left( Q^2-2\,M\,r \right)\tanh^2\XX\,\dd \vv\,\dd\fii \nonumber \\
            & + \frac{2\,a^2\left( Q^2-2\,M\,r \right)\tanh^2\XX}{|\rho|^2\cosh^2\XX\,\cos\teta}\,\dd \teta\,\dd\tilde{\phi}
              - \frac{r^4 + a^2\left( 2\,M\,r+r^2-Q^2 \right)\cosh^{-2}\XX}{|\rho|^2\,\cosh^2\XX\,\cos^2\teta}\dd\teta^2  -
              \frac{\tanh^2\XX}{|\rho|^2}\left( a^2\tilde{\Delta}\sech^{2}\XX+ R \right)\,\dd\tilde{\phi}^2.
  \end{align}
\end{widetext}
This is the metric given in \cite{Fletcher2003}, except for that we used the horizon-penetrating Kerr coordinates instead of Boyer-Lindquist coordinates. The vector
field $n_3^a$ given by \eqref{eq:n congruence} has, in these coordinates, the simple form
\begin{align}\label{eq:n-non-affine}
 n_3 &= - \frac{\sqrt{R}}{|\rho|^2}\,\pd_r,
\end{align}
where $\rho$ is now
\begin{align}\label{eq:rho-bondi}
  \rho &= r + \ii\,a\sech\XX. &                                   
\end{align}
The remaining tetrad vectors read in these coordinates
\begin{widetext}
\begin{subequations}
  \begin{align}
    \ell_3 &= \pd_{\vv} + \frac{\tilde{\Delta}}{P+\sqrt{R}}\,\pd_r  - \frac{a\,\tilde{\Delta}\cos\teta}{\sqrt{R}\,(P+\sqrt{R})}\,\pd_{\teta}  
           + a^{-1}\left(1 - \frac{r^2}{\sqrt{R}} \! \right)\,\pd_{\tilde{\phi}}, \\    
    m_3 &= \frac{1}{\sqrt{2}\,\rho} \!\left[  - \frac{a\,\tilde{\Delta}}{P+\sqrt{R}}\,\pd_r  \
    + \cos\teta\left(\frac{P}{\sqrt{R}}- \ii\,\sinh\XX\right)\pd_{\teta} + \left( \frac{r^2}{\sqrt{R}} + \frac{\ii}{\sinh\XX}
      \right)\pd_{\tilde{\phi}} \right].
                \end{align}\label{eq:l-m}
\end{subequations}
\end{widetext}
The spin coefficient $\tau_3$ has now the form
\begin{subequations}\label{new-coords}
  \begin{align}
    \tau_3 = \frac{a}{\sqrt{2}|\rho|^2} &\left( \frac{\tilde{\Delta}}{P+\sqrt{R}} - 1 + \frac{1}{1 + \ii\,\sinh\XX}\right)\, .\label{eq:new-coords-tau}
  \end{align}
\end{subequations}

\paragraph{Null rotation about $n^a$.} Now we can integrate Eq.\ (\ref{eq:Delta d}) to get
\begin{align}\label{eq:d-def}
  d(r, \teta) &= - \int_{r_+}^r \frac{|\rho(u,\teta)|^2}{\sqrt{R(u)}}\,\tau_3(u,\teta)\,\dd u,
\end{align}
where $\tau_3$ is given by (\ref{eq:new-coords-tau}), and perform null rotation about $n_3^a$, Eq.\ (\ref{np:rotation n}), with the parameter $d$. This yields
the desired Bondi-like tetrad $(\ell_B,n_B,m_B,\bar{m}_B)$. By construction $d$ vanishes on the horizon and hence the triad $(\ell_B,m_B,\bar{m}_B)$ is tangent to
the horizon and parallelly propagated along $n_B^a=n_3^a$.

Since $d$ cannot be evaluated explicitly, we cannot give more explicit expressions for the spin coefficients than (\ref{eq:spin coefficients last}). In terms of those, the spin coefficients for the
rotated tetrad are given by, cf.\ (\ref{np:rotation n spins}),
\begin{subequations}\label{eq:spin-coefficients-final}
  \begin{align}
    \kappa_B &= d(2\,\eps_3+\roo_3)+d^2(\pi_3+2\,\alpha_3)+d^3\,\lambda_3 \nonumber \\
     &\qquad{}+ (\tau_3 + 2\,\beta_3)\,|d|^2 + \mu_3\,d^2\,\bar{d} - |d|^2\Delta_3 d  \\
     &\qquad{}-D_3 d - d\,\bar{\delta}_3d - \bar{d}\,\delta_3 d,\nonumber\\
    \sigma_B &= d(\tau_3+2\,\beta_3)+\mu_3\,d^2 - d\,\Delta_3 d - \delta_3 d, \\
    \roo_B &= \roo_3 + 2\,d\,\alpha_3 +\bar{d}\,\tau_3 + d^2\,\lambda_3 - \bar{d}\,\Delta_3 d - \bar{\delta}_3 d, \\
    \eps_B &= \eps_3 + d(\alpha_3+\pi_3)+\beta_3\,\bar{d} + \lambda_3\,d^2+\mu_3\,|d|^2, \\
    \beta_B &= \beta_3 + d\,\mu_3, \\
    \alpha_B &= \alpha_3 + d\,\lambda_3, \\
    \pi_B &= \pi_3 + d\,\lambda_3 + \bar{d}\,\mu_3, \\
    \mu_B &= \mu_3, \\
    \lambda_B &= \lambda_3, \\
    \tau_B &= \gamma_B = \nu_B = 0,
  \end{align} 
where the spin coefficients with subscript 3 are given by (\ref{eq:spin coefficients last}), the operators $D_3 = l_3^a \nabla_a, \Delta_3 = n_3^a\nabla_a$ and $\delta_3 = m_3^a\nabla_a$ are
given by (\ref{eq:n-non-affine}) and (\ref{eq:l-m}), and the derivatives of $d$ are
\begin{align}
  \frac{\pd d}{\pd r} &= - \frac{|\rho|^2}{\sqrt{R}}\,\tau_3, \\
  \frac{\pd d}{\pd \teta} &= \int_{r_+}^r \frac{a}{\sqrt{2\,R(u)}\,\cos\teta}\,
                            \frac{\ii\,\cosh\XX(u,\teta)}{\left(1+\ii\,\sinh\XX(u,\teta)\right)^2}\,\dd u.
\end{align}
\end{subequations}

In the last step we perform two remaining coordinate transformations. First we define a new radial coordinate by rescaling $r$,
\begin{align}\label{eq:r rescaling}
  \rr &= \int_{r_+}^r \frac{|\rho(u,\teta)|^2}{\sqrt{R(u)}}\,\dd u,
\end{align}
so that $\rr$ is an affine parameter along $n^a$
\begin{align}
  n_B &= - \,\pd_{\rr},
\end{align}
and vanishes on the horizon. In the rest of the paper, the variable $r = r(\rr)$ will always be understood as the function of this new coordinate $\rr$. Next, in order
to eliminate the $\phi$-component of $\ell^a$ on the horizon, we perform the last coordinate transformation
\begin{align}\label{eq:last-phi}
  \phi &= \tilde{\phi} - \frac{a\,\vv}{a^2+r_+^2},
\end{align}
which brings the tetrad into the form (\ref{eq:Krishnan tetrad}). 

\bigskip

\section{Results}
\label{sec:full-tetrad}

\subsection{NP-quantities}

Now we can summarize the obtained results. We have found the Bondi-like coordinates $\xx^\mu=(\vv,\rr,\teta,\phi)$ in which, following the notation of \cite{Krishnan2012}, the Bondi-like null
tetrad is of the form
\begin{align}
  \begin{split}
  \ell_B &= \pd_{\vv} + U\,\pd_{\rr} + X^I\,\pd_I,  \\
  n_B &= -\,\pd_{\rr}, \\
  m_B &= \Omega\,\pd_{\rr} + \xi^I\,\pd_I,
\end{split}\label{eq:full-tetrad}
\end{align}
where $I=2,3$ and the components of the tetrad read
\begin{widetext}
\begin{subequations}
\begin{align}
  U &= \frac{\tilde{\Delta}}{\sqrt{2}\sqrt{R}(P+\sqrt{R})}\left( \sqrt{2}\, |\rho|^2 - a(\rho\,d +\bar{\rho}\,\bar{d})\left(1+\frac{a\cos\teta}{|\rho|^2}\,\frac{\pd \rr}{\pd \teta}\right) -
      \sqrt{2}\,\frac{\pd \rr}{\pd \teta}\,a\cos\teta \right) -|d|^2\nonumber \\
  &\qquad  +{} \sqrt{2}\,\frac{\pd \rr}{\pd \teta} \,\Re\frac{d}{\bar{\rho}}(1+\ii\,\sinh \XX), \\
  X^2 &= \frac{a\,\tilde{\Delta}\cos\teta}{\sqrt{2}\,\sqrt{R}(P+\sqrt{R})}\left(- \sqrt{2} + \frac{a\,\bar{d}}{\rho} + \frac{a\,d}{\bar{\rho}}\right)+
        \frac{\cos \teta}{\sqrt{2}}\left( \frac{\bar{d}}{\rho}\big(1-\ii\,\sinh\XX\big) +\frac{d}{\bar{\rho}}\big(1+\ii\,\sinh\XX\big) \right)\!, \\
  X^3 &=- \frac{a}{a^2+r_+^2}  + a^{-1}\left(1 - \frac{r^2}{\sqrt{R}}\right)
        + \frac{\bar{d}}{\sqrt{2}\,\rho}\left( \frac{r^2}{\sqrt{R}} + \frac{\ii}{\sinh \XX}\right)
        +   \frac{d}{\sqrt{2}\,\bar{\rho}}\left( \frac{r^2}{\sqrt{R}} - \frac{\ii}{\sinh \XX}\right)\!, \\
  \Omega &= - d - \frac{a\,\tilde{\Delta}\,\bar{\rho}}{\sqrt{2}\,\sqrt{R}(P+\sqrt{R})} + \frac{\cos\teta}{\sqrt{2}\,\rho}\frac{\pd \rr}{\pd \teta}\left(\frac{P}{\sqrt{R}}- \ii \sinh\XX\right), \\
  \xi^2 &= \frac{\cos\teta}{\sqrt{2}\,\rho} \bigg( \frac{P}{\sqrt{R}} - \ii\,\sinh\XX\bigg),\\
  \xi^3 &= \frac{1}{\sqrt{2}\,\rho}\!\left( \frac{r^2}{\sqrt{R}} + \frac{\ii}{\sinh\XX}\right)\!,
\end{align}
\end{subequations}
\end{widetext}
and where $\XX$ is given by (\ref{eq:X-def}). Coordinates $\vv, \rr, \teta$ and $\phi$ are related to the standard ingoing null coordinates
$v,r,\theta$ and $\fii$ by (\ref{eq:v-transf}), (\ref{eq:r rescaling}), (\ref{eq:theta-transf}), (\ref{eq:phi-transf}) and (\ref{eq:last-phi}), respectively; the
functions $\alpha$ and $d$ are given by (\ref{eq:alpha-def}) and  (\ref{eq:d-def}), respectively. Functions $\rho,~\tilde{\Delta},~P$ and $R$ are
given by Eqs.\ (\ref{eq:rho Delta def}) and (\ref{eq:Theta P R def}), and the variable $r$ is related to the coordinate $\rr$ by (\ref{eq:r rescaling}). Finally,
\begin{align}
  \frac{\pd \rr}{\pd \teta} &= - \int_{r_+}^r \frac{2\,a^2}{\sqrt{R(u)}}\,\frac{\sinh \XX(u,\teta)}{\cosh^3\XX(u,\teta)\,\cos\teta}\,\dd u.
\end{align}

In the tetrad (\ref{eq:full-tetrad}), the following spin coefficients vanish:
\begin{align}
  \gamma_B &= \nu_B = \tau_B = 0;
\end{align}
these equalities imply that $n_B^a$ is an affinely parametrized geodesic, and both $\ell_B^a$ and $m_B^a$ are covariantly constant along $n_B^a$. The remaining spin coefficients are given
by (\ref{eq:spin-coefficients-final}). In particular, the spin coefficient $\mu_B$ is real which means that the  congruence $n_B^a$ is nontwisting. The spin coefficients $\sigma_B$
and $\lambda_B$ which describe the shear of $\ell_B^a$ and $n_B^a$, respectively, do not vanish; another difference from the Kinnersley tetrad. 

In type D spacetimes and in the tetrad adapted to the principal null directions, the sole nonvanishing Weyl scalar is $\Psi_2$. The Bondi-like
tetrad is not adapted to these null directions anymore so that the full set of Weyl scalars is given by
\begin{subequations}
  \begin{align}
    \Psi^B_0 &= 6\,d^2\left(1 - \frac{a\,d}{\sqrt{2}\,\bar{\rho}}\right)^2\,\Psi_2^K, \\
    \Psi^B_1 &= 3 \,d  \left(\frac{a^2\, d^2}{\bar{\rho }^2}-\frac{3 \,a \,d}{\sqrt{2}\, \bar{\rho
             }}+1\right)\Psi_2^K, \\
    \Psi^B_2 &= \left(\frac{3\, a^2\, d^2}{\bar{\rho }^2}-\frac{3\, \sqrt{2}\, a\, d}{\bar{\rho
             }}+1\right)\Psi_2^K,\\
    \Psi^B_3 &= \frac{3\, a\,}{\bar{\rho}} \left(\frac{a\, d}{\bar{\rho }}-\frac{1}{\sqrt{2}}\right) \Psi_2^K, \\
    \Psi^B_4 &= \frac{3\,a^2}{\bar{\rho}^2} \Psi_2^K,
  \end{align}
\end{subequations}
where $\Psi_2^K$ is given by (\ref{eq:Psi2}). Similarly, for the components of the electromagnetic field we have
\begin{subequations}
  \begin{align}
    \phi^B_0 &= d\left(2- \frac{\sqrt{2}\,a\,d}{\bar{\rho}}\right)\phi_1^K, \\
    \phi^B_1 &= \left(1 - \frac{\sqrt{2}\,a\,d}{\bar{\rho}}\right)\phi_1^K,  \\
    \phi^B_2 &= -\frac{\sqrt{2}\,a}{\bar{\rho}}\,\phi_1^K
  \end{align}
\end{subequations}
where $\phi_1^K$ is given by (\ref{eq:phi1}).

In nonrotating limit, $a = 0$, the tetrad (\ref{eq:full-tetrad}) reduces to corresponding Kinnersley tetrad, since in this case all Lorentz
transformations we applied are identities (except for the spin, whose purpose was to eliminate the coefficient $\gamma$, which vanishes for $a=0$ and, hence, it
is not necessary to perform the spin). In other words, for the Reissner-Nordstr\"om spacetime, the Kinnersley tetrad is already Bondi-like and $\Psi^B_3, \Psi^B_4, \phi^B_0$ and $\phi^B_2$ vanish.

\subsection{Initial data}

Although the tetrad and the corresponding spin coefficients we found are quite lengthy and containing the function $d$ which cannot be integrated explicitly, we have provided
all relations which are necessary to perform calculations in this tetrad. They can be done, for example, symbolically in Mathematica or similar software. The formulas
are also suitable for numerical calculations, since the calculation of the tetrad components and the spin coefficients involves only numerical integration. In this section we use the Bondi-like tetrad to extract appropriate initial data given on the horizon and transversal null hypersurface which reproduce the Kerr-Newman solution, see Sec.\ \ref{sec:WIH}. 


In order to formulate the initial value problem whose solution is the full Kerr-Newman spacetime, we start with the initial data on the initial
sphere $\Sphere_0$. The only nontrivial components of the Bondi-like tetrad on the horizon are
\begin{subequations}
  \begin{align}
    \xi^2 &\horeq  \frac{e^{-\ii\,\teta}}{\sqrt{2}\,\rho\zero}, \\
    \xi^3 &\horeq \frac{1}{\sqrt{2}\,\rho\zero}\!\left(\frac{r_+^2}{P\zero}+\ii\cot\teta\right)\!,
  \end{align}
\end{subequations}
which determine the metric on $\Sphere_0$,
\begin{multline}
  \dd s^2|_{\Sphere_0} = - \left(|\rho\zero|^2 + \frac{a^4\sin^2\teta\cos^2\teta}{|\rho\zero|^2}\right) \dd \teta^2 \\
  -{} \frac{2\,a^2\,(P\zero)^2}{|\rho\zero|^2}\cos\teta\sin^2\teta\,\dd\teta\,\dd\phi - \frac{(P\zero)^2\sin^2\teta}{|\rho\zero|^2}\,\dd\phi^2,
\end{multline}
where the superscript $(0)$ denotes the value of corresponding quantity on $\Sphere^0$. The area element acquires the standard form
\begin{align}
  \dd S &= (r_+^2 + a^2) \sin\teta\,\dd\teta \wedge \dd \phi.
\end{align}
For the spin coefficients we have
\begin{subequations}
  \begin{align}
    \eps\zero &= \frac{r_+-M}{2\,P\zero}, \\
    a\zero &\equiv \alpha\zero-\bar{\beta}\zero = - \frac{r_+ \cos\teta - \ii \,a}{\sqrt{2}\,(\bar{\rho}\zero)^2\,\sin\teta}, \\
    %
      \mu\zero &= \frac{1}{2\,P\zero\,|\rho\zero|^2} \big( -2 \,r_+^3   -a^2(M+r_+) \nonumber \\
        &\qquad\qquad\qquad\qquad{}- a\,P\zero\cos2\teta\,\csc\teta\big),
  \end{align}
  \begin{align}
      \lambda\zero &= \frac{a}{2\,P\zero\,(\bar{\rho}\zero)^3}\Big( - 2 \,a \,Q^2 + a \,r_+(3\,M+r_+) \nonumber \\
        &\qquad\qquad{} + P\zero(r_+\cos 2\teta - \ii\,a\cos\teta)\csc\teta \nonumber \\
        &\qquad\qquad{} + \ii \big(2\,r_+^3+a^2(M+r_+)\big)\cos\teta \Big).
  \end{align}
\label{eq:data on S0}
\end{subequations}
The surface gravity of the horizon is $\surfkappa \horeq 2\,\eps\zero$. To complete the formulation of the initial value problem,
one has to specify the values of $\Psi_4$ and $\phi_2$ on the null hypersurface $\NN_0$ which
intersects the horizon at the initial sphere $\Sphere_0$:
  \begin{align}
    \Psi_4 &= \frac{3\,a^2}{\bar{\rho}^5} \left( - M + \frac{Q^2}{\rho} \right)\!, &
                                                                                   \phi_2 &= - \frac{a\,Q}{\bar{\rho}^3} \quad \text{on}~\NN_0.
  \end{align}

For a general WIH, one has to provide also the values of $\pi\zero$ and $\phi_1\zero$ on $\Sphere_0$, which in our case are
\begin{subequations}
\begin{align}
      \pi\zero &= \frac{a}{\sqrt{2}\,\bar{\rho}\zero}\!\left( \frac{M-r_+}{P\zero}\,e^{-\ii\,\teta} + \frac{\ii}{\bar{\rho}\zero}\sin\teta\right)\!, \\
  \phi_1\zero &= \frac{Q}{\sqrt{2}\,(\bar{\rho}\zero)^2}.
\end{align}
\end{subequations}
However, for a stationary, axially symmetric horizon, these quantities are solutions to the constraints (cf.\ \cite{Lewandowski-Pawlowski-2003})
\begin{subequations}
  \begin{align}
    \bar{\eth}\pi\zero + (\pi\zero)^2 &= \surfkappa\,\lambda\zero,\\
  \delta\zero \phi_1\zero + 2\,\pi\zero \,\phi_1\zero &= \surfkappa\,\phi_2\zero,
\end{align}
\end{subequations}
where $\delta\zero \horeq \xi^I \pd_I$, the operator $\eth$ is defined by (\ref{np:eth}) and $\phi_2\zero$ is the value of $\phi_2$ at $\Sphere_0$. These constraints were the main
ingredients for the proof of the Meissner effect for WIHs in \cite{Guerlebeck-Scholtz-2017}.
The Eqs.\ (\ref{eq:data on S0})
determine $\Psi_2\zero$ and $\Psi_3\zero$ via the Ricci identities (see also \cite{Krishnan2012})
\begin{subequations}
  \begin{align}
    \Re\Psi_2\zero &= |a\zero|^2 - \frac{1}{2}\left( \delta\zero a\zero + \bar{\delta}\zero \bar{a}\zero\right) + |\phi_1\zero|^2, \\
    \Im\Psi_2\zero &= - \Im \eth\pi\zero, \\
    \Psi_3\zero &= (\eth+\bar{\pi}\zero)\lambda\zero - (\bar{\eth}+\pi\zero)\mu\zero.
  \end{align}
\end{subequations}

The evolution of the components of the tetrad (\ref{eq:full-tetrad}) is given by the NP commutators, the evolution of the spin coefficients is governed by the Ricci identities, the evolution of
the Weyl scalars is determined by the Bianchi identities and the evolution of NP components of electromagnetic field is given by the NP form of Maxwell's equations. We do not list these
equations here (see, e.g.\ \cite{Stewart1993}), because we know already that the solution is the Kerr-Newman spacetime in Bondi-like coordinates equipped with the Bondi-like null tetrad constructed
in this paper.

It is worth noting that for a general WIH, $\Psi_4$ and $\phi_2$ are functions given on $\NN_0$ and they are independent from the data given on $\Sphere_0$. This is not the case for the
Kerr-Newman spacetime, since these quantities are directly given by $\Psi_2\zero$ and $\phi_1\zero$ and their transformation properties under the Lorentz transformations. However, as pointed out already in \cite{Lewandowski-2000}, one can vary the data on $\NN_0$ while keeping the initial data on $\Sphere_0$ to produce a wide class of Kerr-like solutions in which the \emph{intrinsic} geometries of the horizon coincide with the geometry of Kerr but differ \emph{off} the horizon.

\section{Conclusions}

The intrinsic properties of the Kerr-Newman black hole are well understood and have been exhaustively investigated in the formalism of (weakly) isolated horizons. Despite the
relatively high degree of symmetry, the Kerr-Newman solution adequately describes isolated black holes in equilibrium in the absence of matter outside the black hole thanks to its uniqueness properties. As we explained in the introduction, the formalism of WIHs allows one to generate a large class of solutions representing black holes deformed by external matter or fields which can be prescribed in an arbitrary way; in particular, the external fields are not restricted to be weak. In order to accomplish this
program and analyze properties of deformed black holes analytically, one first needs the description of the full Kerr-Newman metric not only the part intrinsic to the horizon in
the WIH formalism.

In this paper, we explicitly constructed a Bondi-like tetrad for the Kerr-Newman black hole satisfying the properties imposed in \cite{Krishnan2012} for a general WIH. In this tetrad, we were
able to find the initial data given on the horizon and, more importantly, on the transversal null hypersurface $\NN_0$. In this sense we completed the description of Kerr-Newman solution
in the framework of WIHs. Having the standard example of a WIH at hand, the next step is to consider variations of the initial data and
understand their physical implications. This work is in progress.

\section*{Acknowledgement} 

Work of MS was financially supported by the grant GA\v{C}R 17-16260Y of the Czech Science Foundation. AF acknowledges the support from grant GAUK 588217 of Charles University in Prague.
NG and MS gratefully acknowledge support from the DFG within the Research Training Group
1620 “Models of Gravity”. Partial support of NG comes also
from NewCompStar, COST Action MP1304. AF and MS acknowledge useful discussions with prof.\ Pavel Krtou\v{s} and dr.\ David Kofro\v{n} and the hospitality of institute ZARM, University of Bremen, Germany.

\appendix

\section{Newman--Penrose formalism}
\label{app:np}

For the sake of completeness, in this section we list all relevant definitions and relations of the Newman--Penrose (NP) formalism. In this paper, we follow the
conventions of \cite{Newman-Penrose-1962,PenroseRindlerI,Stewart1993} adapted to the metric signature $(+---)$, while also other conventions are
common \cite{Stephani2009,Griffiths2009}. In particular, our conventions differ from \cite{Krishnan2012}. 

The ``NP tetrad'' is a four-tuple of null vectors $(\ell^a, n^a, m^a, \bar{m}^a)$ normalized by relations
\begin{align}
 \ell^a \, n_a &= - m^a \,\bar{m}_a = 1,
\end{align}
where all other possible contractions vanish. Covariant derivatives in the directions of vectors forming the null tetrad are denoted by
\begin{equation}
\begin{aligned}
 D &= \ell^a \nabla_a, &
 \Delta &= n^a \nabla_a, &
 \delta &= m^a \nabla_a, &
 \bar{\delta} &= \bar{m}^a \nabla_a,
\end{aligned}
\end{equation}
where bar denotes the complex conjugate. In the NP formalism, the connection is encoded in twelve complex spin coefficients defined by
\begin{align}
 \kappa &= m^a D \ell_a, &
  \tau &= m^a \Delta \ell_a, &
  \eps = \frac{1}{2}\left[
n^a D \ell_a - \bar{m}^a D m_a\right],\nonumber \\
\sigma &= m^a \delta \ell_a, &
   \roo& = m^a \bar{\delta} \ell_a,&
   \beta = 
\frac{1}{2}\left[
n^a \delta \ell_a - \bar{m}^a \delta m_a\right], \nonumber\\
\pi &= n^a D \bar{m}_a, &
  \nu &= n^a \Delta \bar{m}_a,&
   \gamma = \frac{1}{2}\left[
n^a \Delta \ell_a - \bar{m}^a \Delta m_a\right],  \nonumber
 \\
 \lambda &= n^a \bar{\delta} \bar{m}_a,&
  \mu &= n^a \delta \bar{m}_a,&
  \alpha = \frac{1}{2}\left[
n^a \bar{\delta} \ell_a - \bar{m}^a \bar{\delta} m_a\right]\,.\nonumber \\
\label{np:spin coeffs def}
\end{align}
Some of the spin coefficients have direct geometrical meaning \cite{Krishnan2012}. Namely, real and imaginary parts of $\roo$ determine the expansion and twist \cite{Poisson2004} of the congruence $\ell^a$, respectively; similarly, real and imaginary parts of $\mu$ describe the expansion and twist of $n^a$ \cite{Krishnan2014}; coefficients $\sigma$ and $\lambda$ describe the shear of $\ell^a$ and $n^a$. Definitions \eqref{np:spin coeffs def} imply ``transport equations''
\begin{subequations}
\begin{align}
  D \ell^a &= \left( \eps+\bar{\eps} \right)\ell^a - \bar{\kappa}\,m^a - \kappa\,\bar{m}^a, & \\
  \Delta n^a &= -\left( \gamma+\bar{\gamma} \right)n^a  + \nu\,m^a + \bar{\nu}\,\bar{m}^a, \\
  \Delta \ell^a &= (\gamma+\bar{\gamma})\ell^a - \bar{\tau}\,m^a - \tau\,\bar{m}^a,
  \label{np:transport D l Delta n}
\end{align}
\end{subequations}
which show that $\kappa$ and $\nu$ describe the deviation of $\ell^a$ and $n^a$ from being geodesics, while $\eps+\bar{\eps}$ and $\gamma+\bar{\gamma}$ measure the failure of $\ell^a$
and $n^a$ being affinely parametrized. Covariant derivative of $m^a$ along $n^a$ is given by the transport equation
\begin{align}
 \Delta m^a &= \bar{\nu}\ell^a - \tau\,n^a + \left( \gamma-\bar{\gamma} \right)m^a.\label{np:transport Delta m}
\end{align}
Components of the Weyl tensor with respect to the null tetrad are provided by the Weyl scalars
\begin{align}
\Psi_0 &= C_{abcd}\ell^a m^b \ell^c m^d, &
\Psi_1 &= C_{abcd}\ell^a n^b \ell^c m^d, &\nonumber\\
\Psi_2 &= C_{abcd}\ell^a m^b\bar{m}^c n^d,&
\Psi_3 &= C_{abcd}\ell^a n^b\bar{m}^c n^d,\nonumber\\
\Psi_4 &= C_{abcd}\bar{m}^a n^b\bar{m}^cn^d. \label{np:Weyl scalars}
\end{align}
In electrovacuum spacetimes, the components of the trace-free part of the Ricci tensor are given by $\Phi_{mn} = \phi_m\,\bar{\phi}_n$, where
\begin{equation}
\begin{aligned}
\phi_0 &=  F_{ab} \ell^a m^b,  \qquad 
\phi_2 =  F_{ab} \bar{m}^a n^b, \\
\phi_1 &=  \frac{1}{2} F_{ab} \left[\ell^a n^b  -  m^a 
\bar{m}^b\right]\!. &
\label{np:EM components}
\end{aligned}
\end{equation}
are the components of the electromagnetic tensor $F_{ab}$. The scalar curvature is $\Lambda = 0$. 

The actual field equations are provided by the set of Ricci identities, Bianchi identities and commutation relations. The full list of these equations as well
as the Maxwell equations in the NP formalism can be found, e.g., in \cite{PenroseRindlerI,Stewart1993}. In the present paper, we need only few of those
equations and, hence, we show them in the appropriate context only.

On the other hand, we employ all freedom available in the choice of the tetrad and for that we need the complete set of transformation equations for the
NP quantities. A ``boost'' in the plane spanned by $\ell^a$ and $n^a$ with real parameter $A$ is defined as the transformation
\begin{align}
 \ell^a &\mapsto A^2\,\ell^a, &
 n^a &\mapsto A^{-2}\,n^a, &
 m^a & \mapsto m^a.
 \label{np:boost}
\end{align}
Under boost, the spin coefficients \eqref{np:spin coeffs def} transform accor\-ding to the formulae
\begin{align}
 \kappa &\mapsto A^4\,\kappa, &
  {\tau} &\mapsto \tau, & 
  {\sigma} &\mapsto A^2\,\sigma, &
  {\roo} &\mapsto A^2\,\roo, & \nonumber \\
  {\pi} &\mapsto \pi, &
  {\nu} &\mapsto A^{-4}\,\nu,&
  {\mu} &\mapsto A^{-2}\,\mu,&
  {\lambda} &\mapsto A^{-2}\,\lambda. \nonumber
\end{align}
\begin{align}
 {\eps} &\mapsto A^2\,\eps + A\,D A, &
 {\gamma} &\mapsto A^{-2}\,\gamma + A^{-3}\,\Delta A, & \nonumber \\
 {\beta} &\mapsto \beta + A^{-1}\,\delta A, &
 {\alpha} &\mapsto \alpha + A^{-1}\,\bar{\delta} A, & \label{np:boost spins}
\end{align}
while the Weyl scalars \eqref{np:Weyl scalars} and electromagnetic scalars \eqref{np:EM components} transform as
\begin{subequations}
  \begin{align}
    \Psi_m &\mapsto A^{2(2-m)}\,\Psi_m, & m&=0,1,2,3,4, \label{np:boost weyls}\\    
    \phi_m &\mapsto A^{2(1-m)}\,\phi_m, & m&=0,1,2. \label{np:boost EM}
  \end{align}
\end{subequations}
Any quantity $\eta$ which transforms as $\eta \mapsto A^{2 w} \eta$ is said to have a ``boost weight'' $w$. 

The next transformation is the spin in the space-like plane spanned by $m^a$ and $\bar{m}^a$ with a real parameter $\chi$ defined by
\begin{align}
 \hat{\ell}^a & \mapsto \ell^a, &
 \hat{n}^a &\mapsto n^a, &
 \hat{m}^a & \mapsto e^{2\,\ii\,\chi}m^a.
 \label{np:spin}
\end{align}
Under spin, the spin coefficients \eqref{np:spin coeffs def} transform as
\begin{align}
 \kappa & \mapsto e^{2\,\ii\,\chi} \kappa, & \tau &\mapsto e^{2\,\ii\,\chi}\tau, &
 \sigma &\mapsto e^{4\,\ii\,\chi} \sigma, & \roo &\mapsto \roo, \nonumber\\
 \pi &\mapsto e^{-2\,\ii\,\chi}\pi, & \nu &\mapsto e^{-2\,\ii\,\chi}\nu, &
 \mu &\mapsto \mu, & \lambda &\mapsto e^{-4\,\ii\,\chi} \lambda,\nonumber
\end{align}
\begin{align}
 \eps & \mapsto \eps + \ii D\chi, & \gamma & \mapsto \gamma + \ii \Delta \chi, & \nonumber \\
 \beta & \mapsto e^{2\,\ii\,\chi}\left( \beta+\ii\delta\chi \right), &
 \alpha &\mapsto e^{-2\,\ii\,\chi}\left( \alpha+\ii\bar{\delta}\chi \right),\label{np:spin spins}
\end{align}
The Weyl scalars \eqref{np:Weyl scalars} as well as the electromagnetic scalars \eqref{np:EM components} are then given by
\begin{align}
\Psi_m &\mapsto e^{2(2-m)\ii\,\chi}\,\Psi_m, & m&=0,1,2,3,4, \label{np:spin weyls}\\
\phi_m &\mapsto e^{2(1-m)\ii\,\chi}\,\phi_m, & m&=0,1,2. \label{np:spin EM}
\end{align}
Again, a quantity $\eta$ is said to have a ``spin weight'' $s$ if it transforms like $\eta \mapsto e^{2\,\ii\,s\,\chi}$ under the spin.
The associated spin raising/lowering operators $\eth$ and $\bar{\eth}$ are defined by \cite{Goldberg1967,Stewart1993}
\begin{align}\label{np:eth}
 \eth\eta &= \delta \eta + s \left( \bar{\alpha}-\beta \right)\eta, &
 \bar{\eth}\eta &= \bar{\delta}\eta - s \left( \alpha - \bar{\beta} \right)\eta.
\end{align}

Another kind of transformation of the null tetrad is a ``null rotation'' about $\ell^a$ with complex parameter $c$:
\begin{align}\label{np:rotation l}
\begin{split}
 \ell^a & \mapsto \ell^a, \qquad m^a \mapsto m^a + \bar{c}\,\ell^a, \\
 n^a &\mapsto n^a + c\,m^a + \bar{c}\,\bar{m}^a + |c|^2\,\ell^a,  
\end{split}
\end{align}
under which the spin coefficients \eqref{np:spin coeffs def} transform as follows:
\begin{align}
 {\kappa} &\mapsto \kappa, \nonumber\\
 {\tau} &\mapsto \tau + c\,\sigma + \bar{c}\,\roo + \kappa\,|c|^2, \nonumber\\
 {\sigma} &\mapsto \sigma + \kappa\,\bar{c}, \nonumber\\
 {\roo} &\mapsto \roo + \kappa\,c,\nonumber\\
 {\eps} &\mapsto \eps + c\,\kappa, \nonumber\\
 {\gamma} &\mapsto \gamma + c(\beta+\tau)+\alpha\,\bar{c} + \sigma\,c^2 + \left( \eps+\roo \right)|c|^2 + \kappa\,c^2\,\bar{c},\nonumber\\
 {\beta} &\mapsto \beta + c\,\sigma + \eps\,\bar{c} + \kappa\,|c|^2, \nonumber\\
 {\alpha} &\mapsto \alpha + c\left( \eps+\roo \right)+\kappa\,c^2, \nonumber\\
 {\pi} &\mapsto \pi + 2\,c\,\eps + c^2\,\kappa + Dc, \nonumber\\
 {\nu} &\mapsto \nu + c\left( 2\gamma+\mu \right) + \bar{c}\,\lambda + c^2\left( 2\,\beta+\tau \right) + c^3\,\sigma  \nonumber\\ 
 &\qquad  + |c|^2 \left( \pi+2\,\alpha \right) + c^2\,\bar{c}\left( 2\eps+\roo \right)+c^3\,\bar{c}\,\kappa  \nonumber \\ 
 &  \qquad + 
  |c|^2 \,Dc + \Delta c + c\,\delta c + \bar{c} \,\bar{\delta} c,  \nonumber\\
  {\mu} &\mapsto \mu + 2\,c\,\beta + \bar{c}\,\pi + c^2\,\sigma + 2\,|c|^2\,\eps + c^2\,\bar{c}\,\kappa + \bar{c}\,Dc + \delta c, \nonumber\\
  {\lambda} &\mapsto \lambda + c\left( \pi + 2\,\alpha \right)+c^2\left( \roo + 2\,\eps \right) + \kappa\,c^3 + c\,Dc +\bar{\delta} c.
  \label{np:rotation l spins}
\end{align}
The transformation rules for the Weyl and electromagnetic scalars \eqref{np:Weyl scalars} and \eqref{np:EM components} read
\begin{align}
 \Psi_0 &\mapsto \Psi_0, \nonumber\\
 \Psi_1 &\mapsto \Psi_1 + c\,\Psi_0, & \nonumber\\
 \Psi_2 &\mapsto \Psi_2 + 2\,c\,\Psi_1 + c^2\,\Psi_0, & \nonumber\\
 \Psi_3 &\mapsto \Psi_3 + 3\,c\,\Psi_2 + 3\,c^2\,\Psi_1 + c^3\,\Psi_0,& \nonumber\\
 \Psi_4 &\mapsto \Psi_4 + 4\,c\,\Psi_3 + 6\,c^2\,\Psi_2 + 4\,c^3\,\Psi_1 + c^4\,\Psi_0,\nonumber\\
 \phi_0 &\mapsto \phi_0, \nonumber \\
 \phi_1 &\mapsto \phi_1 + c\,\phi_0, \nonumber \\
 \phi_2 &\mapsto \phi_2 + 2\,c\,\phi_1 + c^2\,\phi_0.\label{np:rotation l weyls EM}
\end{align}

Finally, a ``null rotation'' about $n^a$ with complex parameter $d$ is defined by the relations
\begin{align}\label{np:rotation n}
\begin{split}
  n^a &\mapsto{n}^a, \qquad m^a \mapsto {m}^a + d\,{n}^a, \\
  \ell^a &\mapsto{\ell}^a + \bar{d}\,{m}^a + d\,{\bar{m}}^a + |d|^2 \,{n}^a.
\end{split}
\end{align}
The spin coefficients \eqref{np:spin coeffs def} now transform as
\begin{align}
 \kappa & \mapsto \kappa + d\left( 2\eps+\roo \right) + \bar{d}\,\sigma + d^2\left( \pi+2\alpha \right) + d^3\,\lambda \,+ \nonumber \\
 &\qquad +|d|^2\left( \tau+2\beta \right)+d^2\,\bar{d}\left( 2\gamma+\mu \right)+d^3\,\bar{d}\,\nu \,- \nonumber \\
 &\qquad - |d|^2\Delta d - D d - d\,\bar{\delta}d - \bar{d}\delta d, \nonumber \\
 \tau &\mapsto \tau + 2\,d\,\gamma + d^2\,\nu-\Delta d, \nonumber \\
 \sigma &\mapsto \sigma + d\left( \tau+2\beta \right)+d^2\left( \mu+2\gamma \right)+d^3\,\nu-d\Delta d-\delta d, \nonumber \\
 \roo &\mapsto \roo + 2\,d\,\alpha + \bar{d}\,\tau + d^2\,\lambda + 2|d|^2\,\gamma + d^2\,\bar{d}\,\nu - \bar{d}\Delta d - \bar{\delta}d,\nonumber\\
 \eps &\mapsto \eps+d\left( \alpha+\pi \right)+\beta\,\bar{d} + \lambda\,d^2 + \left( \mu+\gamma \right)|d|^2+\nu\,d^2\,\bar{d}, \nonumber \\
 \gamma &\mapsto \gamma + d\,\nu, \nonumber \\
\beta &\mapsto \beta + d\left( \gamma+\mu \right)+d^2\,\nu, \nonumber \\
\alpha &\mapsto \alpha + d\,\lambda + \bar{d}\,\gamma + |d|^2\,\nu, \nonumber \\
\pi &\mapsto \pi + d\,\lambda + \bar{d}\,\mu + |d|^2\,\nu, \nonumber \\
\nu & \mapsto \nu, \nonumber \\
\mu &\mapsto \mu + d\,\nu, \nonumber \\
\lambda &\mapsto \lambda + \nu\,\bar{d}.\label{np:rotation n spins}
\end{align}
For the Weyl scalars \eqref{np:Weyl scalars} and the electromagnetic scalars \eqref{np:EM components} we now have
\begin{align}
 \Psi_0 & \mapsto \Psi_0 + 4\,d\,\Psi_1 + 6\,d^2\,\Psi_2+4\,d^3\,\Psi_3 + d^4\,\Psi_4, \nonumber \\
 \Psi_1 &\mapsto \Psi_1 + 3\,d\,\Psi_2 + 3\,d^2\,\Psi_3+d^3\,\Psi_4, \nonumber\\
 \Psi_2 &\mapsto \Psi_2 + 2\,d\,\Psi_3 + d^2\,\Psi_4, \nonumber \\
 \Psi_3 &\mapsto \Psi_3 + d\,\Psi_4, \nonumber \\
 \Psi_4 &\mapsto \Psi_4, \nonumber \\
 \phi_0 &\mapsto \phi_0 + 2\,d\,\phi_1 + d^2\,\phi_2, \nonumber \\
 \phi_1 &\mapsto \phi_1 +d\,\phi_2, \nonumber \\
 \phi_2 &\mapsto \phi_2. \label{np:rotation n Weyls EM}
\end{align}

\section{Visualization}
\label{sec:visualization}

In this section, we briefly present the visualization of the differences between the standard twisting Kinnersley tetrad and the nontwisting tetrad
constructed in this paper. Standard treatment of optical scalars for null geodesics in the NP formalism can be found, e.g., in \cite{Stewart1993}. Here we need to consider
slightly more general case.

Consider a general NP tetrad $\ell^a$, $n^a$ and $m^a$, for which $\ell^a$ is not necessarily a geodesic (which is the case
of $\ell^a_B$ given by \ref{sec:full-tetrad}) and $m^a$ is not necessarily parallelly propagated along either $\ell^a$ or $n^a$ (which is the case
for the Kinnersley tetrad (\ref{eq:kinnersley})). We can always introduce complex null vectors $\xi^a_\ell$ and $\xi_n^a$  for which $D \xi^a_\ell = 0$ and
$\Delta \xi^a_n = 0$, respectively. Let $z^a_\ell$ and $z^a_n$ be deviation vectors orthogonal to $\ell^a$ and $n^a$, respectively, which are
propagated by equations
\begin{align}
  D z_\ell^a &= z_\ell^b\nabla_b \ell^a, &
                                           \Delta z^a_n &= z_n^b\nabla_b n^a.                                           
\end{align}
We expand both connecting vectors as
\begin{align}
  \begin{split}
    z^a_\ell &= c_\ell\,\ell^a - \bar{z}_\ell\,\xi_\ell^a - z_\ell\,\bar{\xi}_\ell^a, \\
    z^a_n &= c_n\, n^a - z_n\,\xi_n^a - \bar{z}_n\,\bar{\xi}_n^a,    
  \end{split}
\end{align}
and interpret the component $z_\ell = x_\ell+\ii\,y_\ell$ ($z_n=x_n+\ii\,y_n$) as complex coordinate of the projection of $z^a_\ell$ ($z^a_n$) onto the spacelike plane orthogonal to $\ell^a$ ($n^a$). Corresponding
evolution equations read
\begin{align}
  \begin{split}
    D z_\ell &= - \roo\,z_\ell-\sigma\,\bar{z}_\ell + \kappa\,c_\ell,  \\
    \Delta z_n &= \mu\,z_n + \lambda\,\bar{z}_n - \nu\,c_n,
  \end{split}\label{eq:DDeltaz}
\end{align}
and
\begin{align}
  \begin{split}
    D c_\ell &= (\pi-\alpha-\bar{\beta})z + (\bar{\pi}-\bar{\alpha}-\beta)\bar{z},   \\
    \Delta c_n &= (\bar{\alpha}+\beta-\tau)z+(\alpha-\bar{\beta}-\bar{\tau})\bar{z}.
  \end{split}
\end{align}

In what follows we consider vectors $\ell^a_K$ and $n^a_K$ of the Kinnersley tetrad and $\ell^a_B$ and $n_B^a$ of the Bondi-like tetrad. In each case we choose
the initial spacetime point with coordinates $(\rr_0,\teta_0)$ and a sequence $\{z_{0,n}=e^{2\pi\,\ii\,n/N}\}_{n=0}^N$ of initial coordinates of the deviation vector. For each
$z_{0,n}$ we solve the deviation equations (\ref{eq:DDeltaz}) along the null geodesic and plot points $z_n$ for given values of the parameter along the geodesic, which we
interpret as the cross-sections of the family of nearby geodesics with initially circular cross-section. 

\begin{figure*}
    \begin{subfigure}{0.4\textwidth}
      \includegraphics[width=0.9\textwidth]{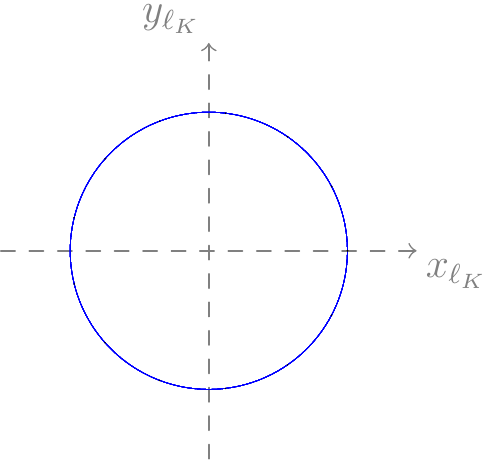}                                                 
    \end{subfigure}
    \begin{subfigure}{0.3\textwidth}
       \includegraphics[width=0.9\textwidth]{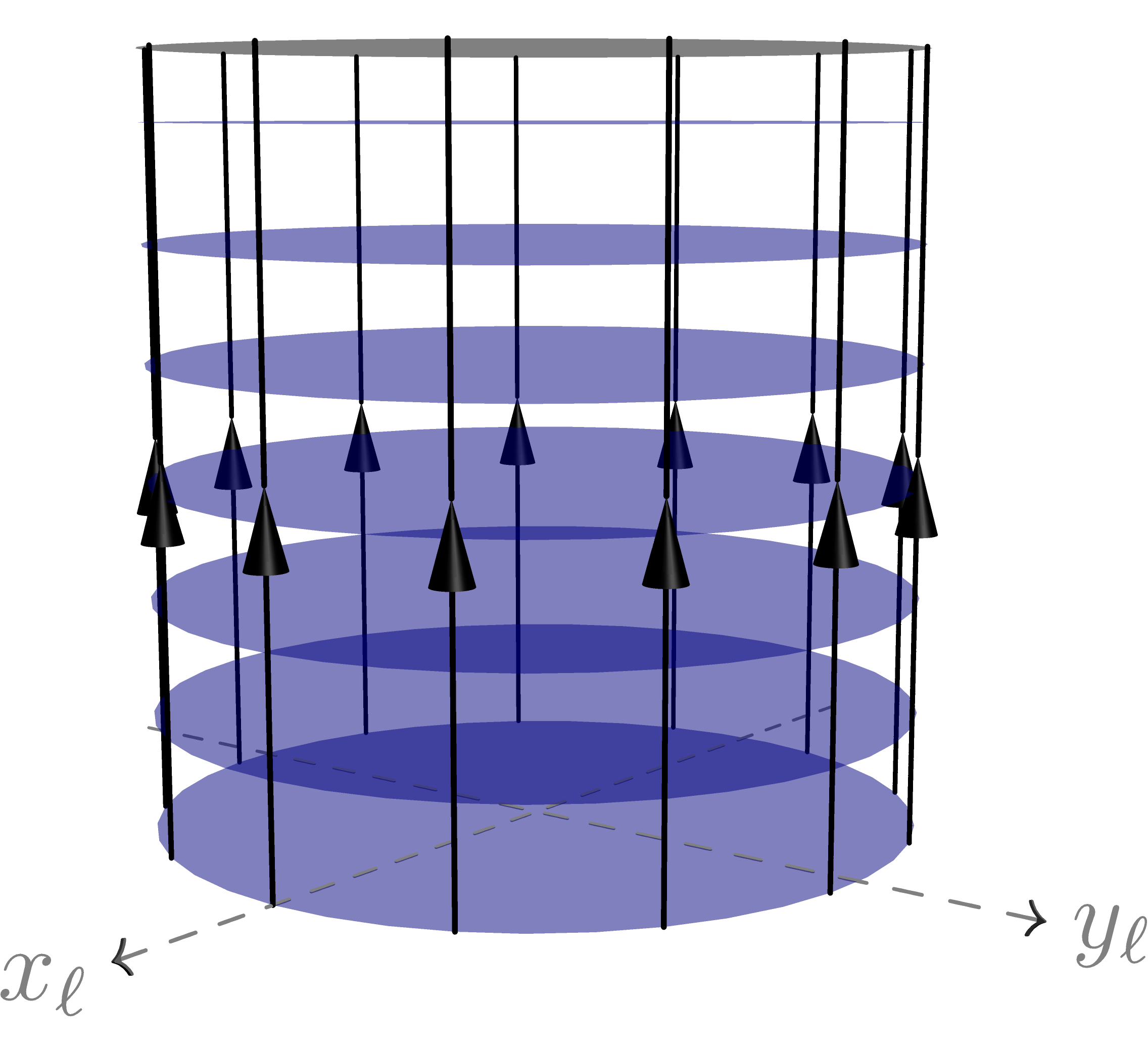}
    \end{subfigure}
    \caption{Nontwisting, nonexpanding and shear-free congruence $\ell_K^a$  with the initial position $\rr_0 = 0$, $\teta_0 = \pi/4$. }
    \label{fig:lK-1}
  \end{figure*}

\begin{figure*}
    \begin{subfigure}{0.4\textwidth}
      \includegraphics[width=0.9\textwidth]{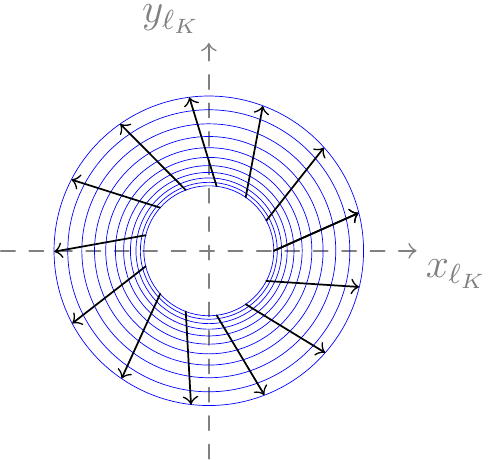}                                                 
    \end{subfigure}
    \begin{subfigure}{0.3\textwidth}
       \includegraphics[width=0.9\textwidth]{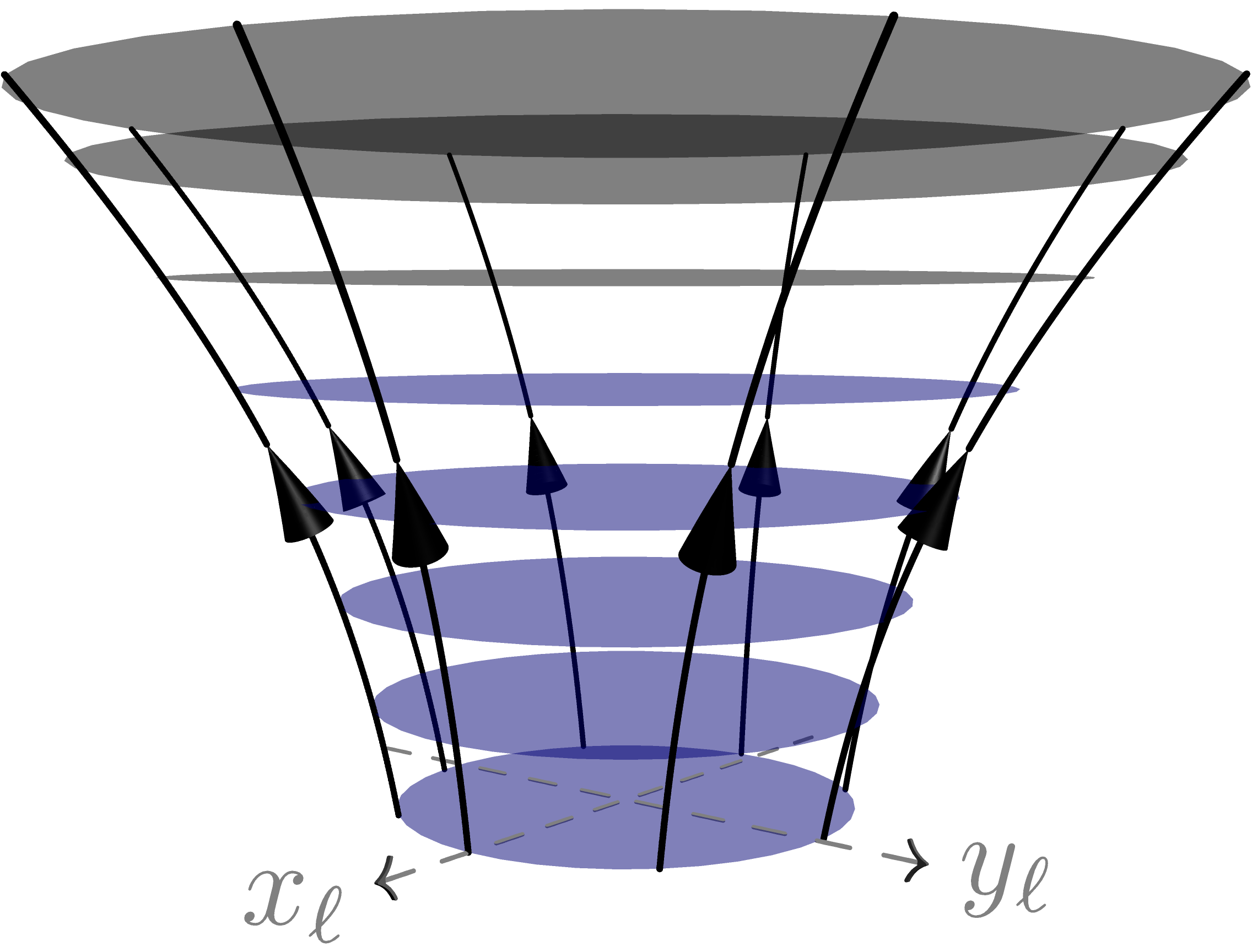}
    \end{subfigure}
    \caption{Twisting, expanding and shear-free congruence $\ell_K^a$ with the initial position $\rr_0 = 0.5$, $\teta_0 = \pi/4$. }
    \label{fig:lK-2}
  \end{figure*}

\begin{figure*}
    \begin{subfigure}{0.4\textwidth}
      \includegraphics[width=0.9\textwidth]{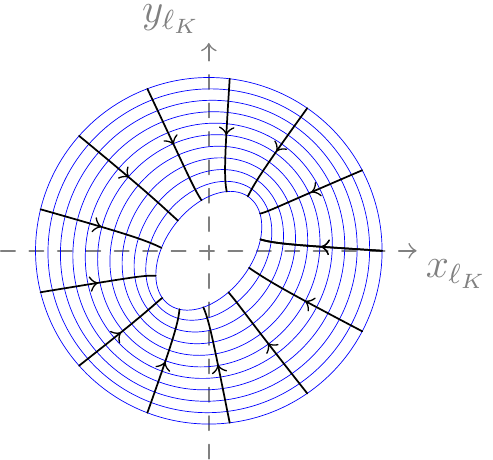}                                                 
    \end{subfigure}
    \begin{subfigure}{0.3\textwidth}
       \includegraphics[width=0.9\textwidth]{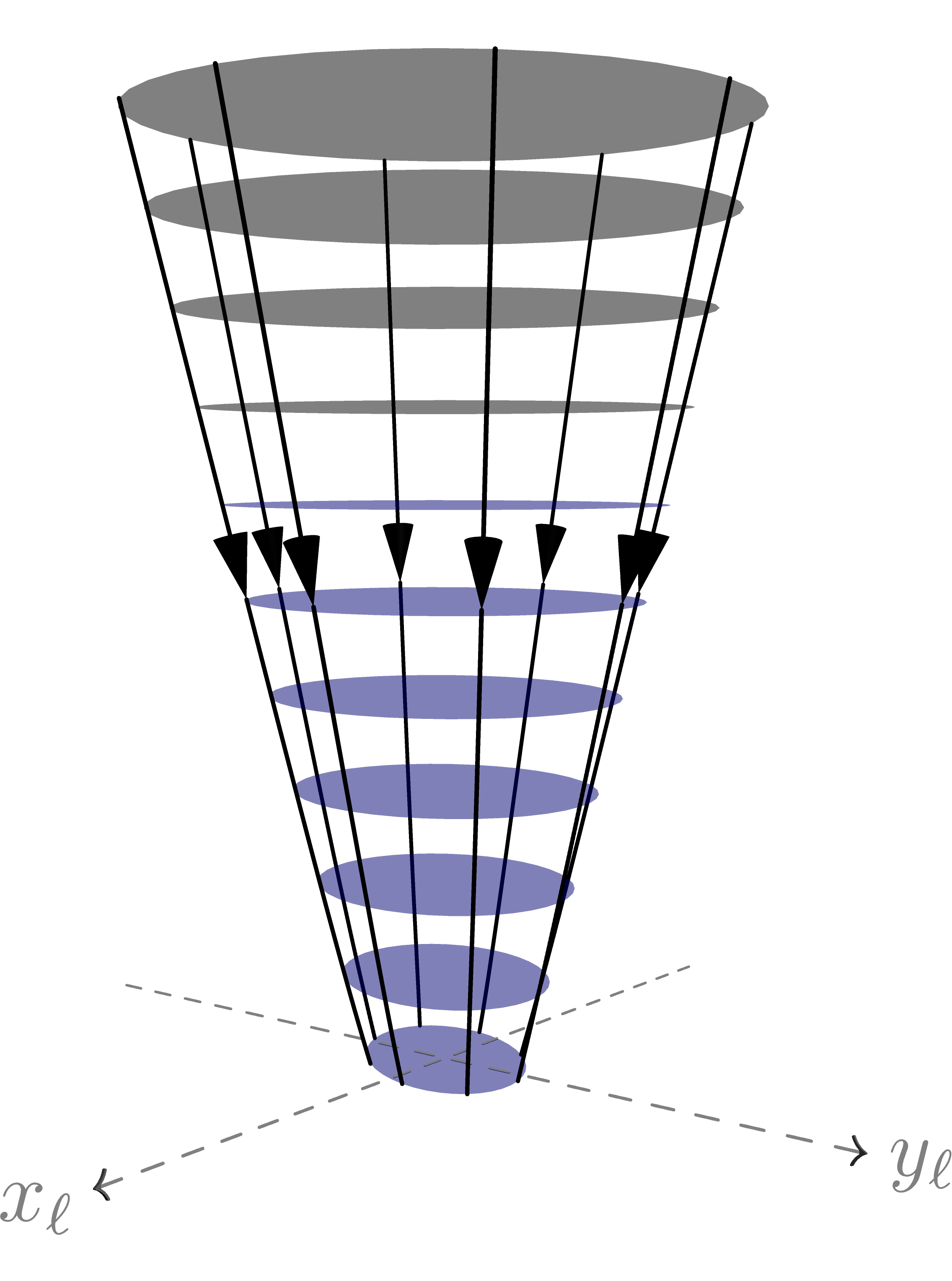}
    \end{subfigure}
    \caption{Nontwisting, converging and shearing congruence $n_B^a$  for the initial position $\rr_0=5$, $\teta_0 = \pi/4$. }
    \label{fig:n}
  \end{figure*}

\begin{figure*}
    \begin{subfigure}{0.4\textwidth}
      \includegraphics[width=0.9\textwidth]{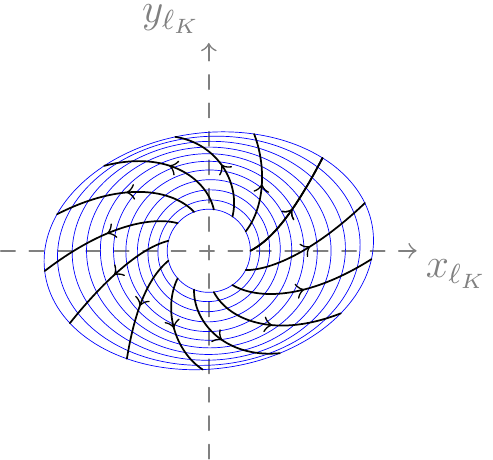}                                                 
    \end{subfigure}
    \begin{subfigure}{0.3\textwidth}
       \includegraphics[width=0.9\textwidth]{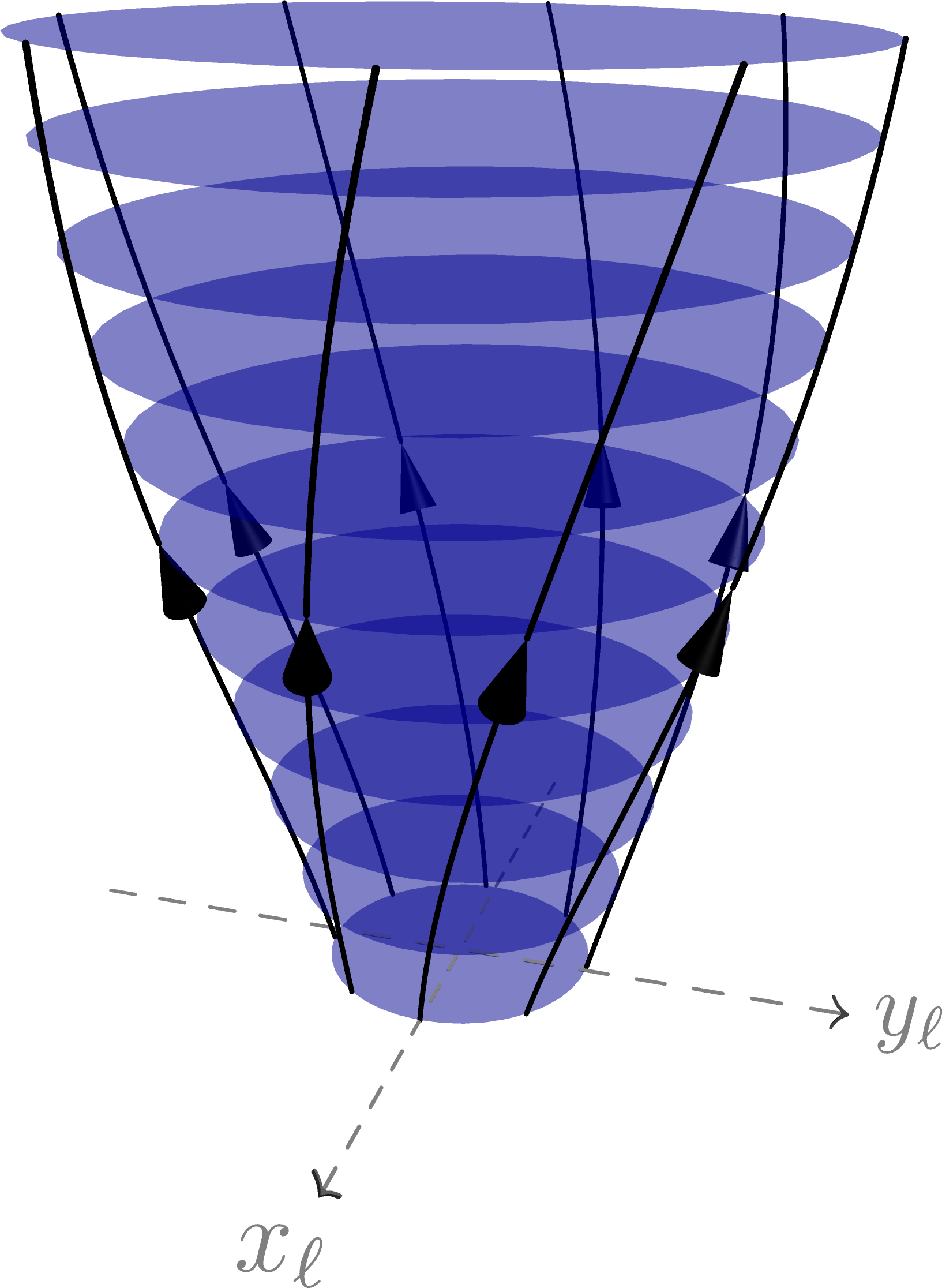}
    \end{subfigure}
    \caption{Twisting, expanding and shearing congruence $\ell_K^a$ for the initial position $\rr_0=2$, $\teta_0 = \pi/4$. }
    \label{fig:l}
  \end{figure*}

We choose the parameters of the Kerr-Newman spacetime $M=1.2, a = 1.1, Q = 0.2$. In Fig.\ \ref{fig:lK-1} we plot the congruence $\ell^a_K \horeq \ell^a_B$ on the horizon. By the properties
of a WIH, this congruence is nontwisting, nonexpanding and shear-free. Congruence $\ell^a_K$ off the horizon, see Fig.\ \ref{fig:lK-2}, is expanding, twisting and shear-free, i.e.\
the cross-sections remain circular. On the other hand, congruence $n_K^a$, being transversal to the horizon and future pointing, is converging and twisting, see Fig.\ {fig:nK}. Congruence
$n_B^a$ of the Bondi-like tetrad is also converging but has zero twist and nonvanishing shear, as Fig.\ \ref{fig:n} demonstrates. Finally, optical scalars of the congruence $\ell^a_B$
vanish on the horizon, Fig.\ \ref{fig:lK-1}, but off the horizon all optical scalars are nonvanishing, see Fig.\ \ref{fig:l}.

  \begin{figure*}
    \begin{subfigure}{0.4\textwidth}
      \includegraphics[width=0.9\textwidth]{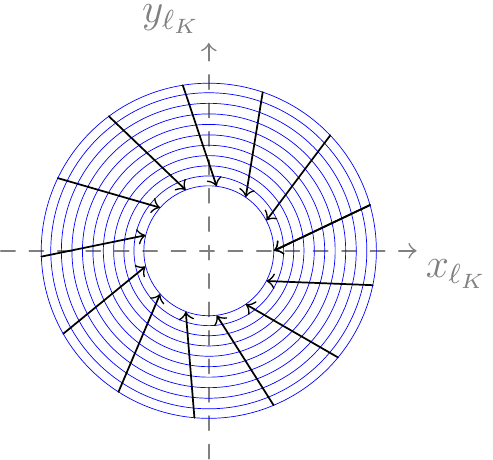}                                                 
    \end{subfigure}
    \begin{subfigure}{0.3\textwidth}
       \includegraphics[width=0.9\textwidth]{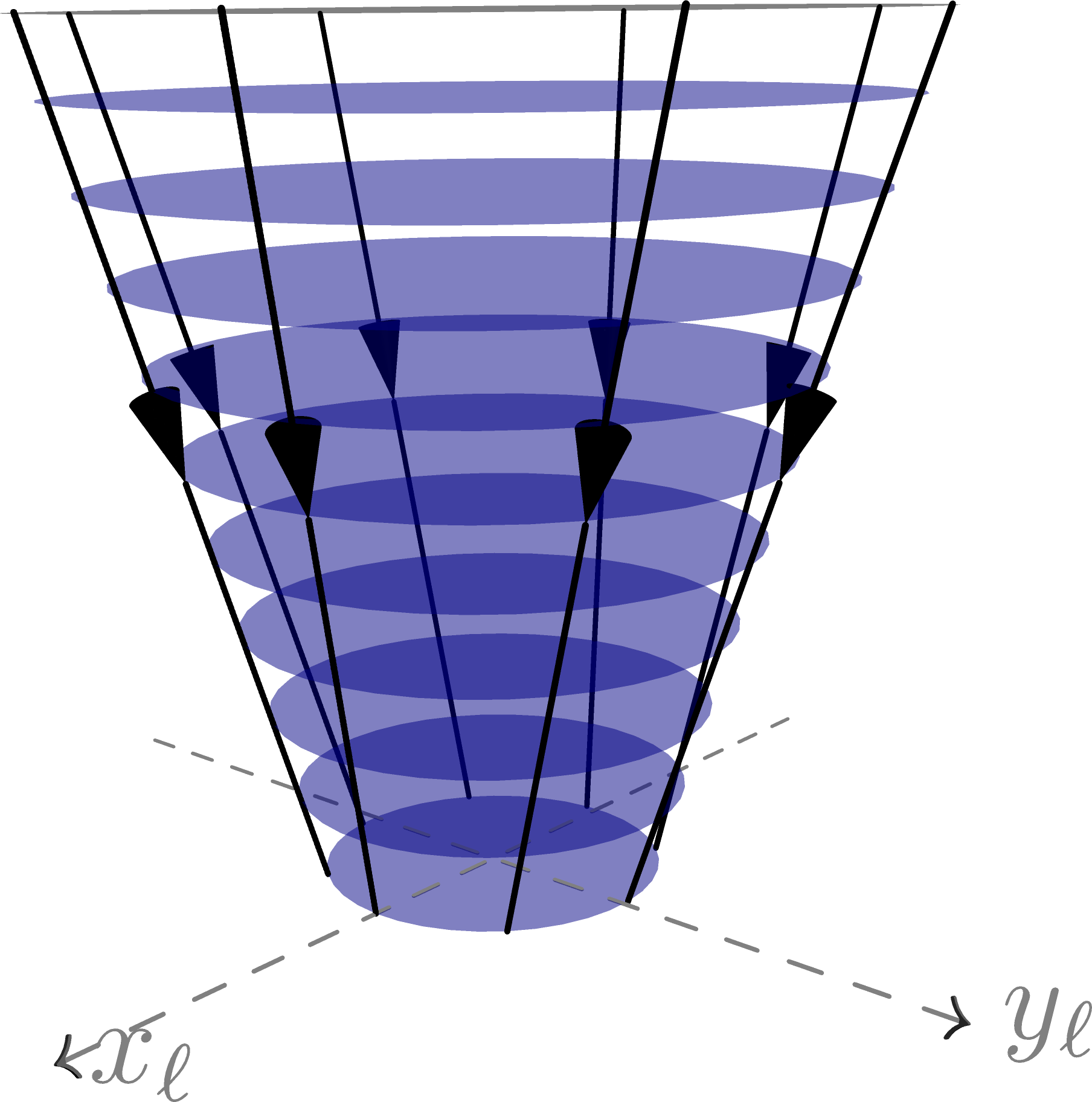}
    \end{subfigure}
    \caption{Twisting, converging and shear-free congruence $n_K^a$ for the initial position $\rr_0 = 3$, $\teta_0 = \pi/4$. }
    \label{fig:nK}
  \end{figure*}

\section*{References}

%

\end{document}